\documentclass[prx,twocolumn,amssymb,superscriptaddress]{revtex4-2}
\usepackage{natbib,amsmath}
\usepackage{graphicx}
\usepackage{multirow}
\usepackage{mathtools}
\usepackage{mathptmx}
\usepackage{fouridx}
\usepackage{amsthm}
\usepackage{eucal}
\usepackage{bm}
\usepackage{upgreek}
\usepackage{framed}
\usepackage{mathrsfs}
\usepackage{latexsym}
\usepackage{float}
\usepackage{subfigure}
\usepackage[dvipsnames]{xcolor}
\usepackage{hyperref,tikz}
\usepackage[utf8]{inputenc}
\usepackage{bbold}
\usepackage{bbm}
\usetikzlibrary{decorations.pathmorphing, patterns,shapes}

\newcommand{\be}{\begin{equation}}
	\newcommand{\ee}{\end{equation}}
\newcommand{\bea}{\begin{eqnarray}}
	\newcommand{\eea}{\end{eqnarray}}
\newcommand{\ba}{\begin{array}}
	\newcommand{\ea}{\end{array}}

\newcommand{\ket}[1]{|#1\rangle}
\newcommand{\bra}[1]{\langle #1|}
\newcommand{\inner}[2]{\langle#1|#2\rangle}

\newcommand{\D}{\mathrm{d}}

\newcommand{\E}[1]{\mathrm{e}^{\mbox{\footnotesize$#1$}}}
\newcommand{\ERF}[1]{\mathrm{erf}\!\left(#1\right)}
\newcommand{\LambertW}[1]{\mathrm{W}\!\left(#1\right)}

\begin{document}	

\title{All-photonic architecture for scalable quantum computing with Greenberger-Horne-Zeilinger states}
	
\author{Srikrishna Omkar}
\email{omkar.shrm@gmail.com}
\affiliation{Department of Physics and Astronomy, Seoul National University, 08826 Seoul, Korea}
\author{Seok-Hyung Lee}
\affiliation{Department of Physics and Astronomy, Seoul National University, 08826 Seoul, Korea}
\author{Yong Siah Teo}
\affiliation{Department of Physics and Astronomy, Seoul National University, 08826 Seoul, Korea}
\author{Seung-Woo Lee}\affiliation{Center for Quantum Information, Korea Institute of Science and Technology, Seoul, 02792, Korea}
\author{Hyunseok Jeong}
\email{h.jeong37@gmail.com}
\affiliation{Department of Physics and Astronomy, Seoul National University, 08826 Seoul, Korea}
	
\begin{abstract}
Linear optical quantum computing is beset by the lack of deterministic entangling operations besides photon loss. Motivated by advancements at the experimental front in deterministic generation of various kinds of multiphoton entangled states, we propose an architecture for linear-optical quantum computing that harnesses the availability of three-photon Greenberger-Horne-Zeilinger (GHZ) states. Our architecture and its subvariants use polarized photons in GHZ states, polarization beam splitters, delay lines, optical switches and on-off detectors. We concatenate topological quantum error correction code with three-qubit repetition codes and estimate that our architecture can tolerate remarkably high photon-loss rate of $11.5 \%$; this makes a drastic change that is at least an order higher than those of  known proposals. Further, considering three-photon GHZ states as resources, we estimate the resource overheads to perform gate operations with an accuracy of $10^{-6}$~($10^{-15}$) to be $2.07\times10^6~(5.03\times10^7)$. Compared to other all-photonic schemes, our architecture is also resource-efficient. In addition, the resource overhead can be even further improved if larger GHZ states are available. Given its striking enhancement in the photon loss threshold and the recent progress in generating multiphoton entanglement, our scheme will make scalable photonic quantum computing a step closer to reality.
\end{abstract}
	
\maketitle

\section{Introduction}

Out of the many physical platforms available for quantum computing, optical platforms facilitate quicker gate operations compared to the decoherence time, fast readouts and efficient qubit transfer~\cite{RevMod07,RP10}. These features make them one of the strongest contenders for realizing scalable quantum computing. Linear optical quantum computing~\cite{RevMod07,KLM01,J07} uses only beam splitters, phase shifters and photon detectors to process the quantum information encoded in light. Besides this apparent simplicity, the ability to operate at room temperature makes this approach attractive. Measurement-based quantum-computing~\cite{RB01} is particularly suitable for optical platforms~\cite{Niel04} in terms of practical implementation. In this approach, a particular class of multi-qubit entangled states, known as cluster states, are first generated, and single-qubit-measurements are then performed on them to realize a universal set of gate operations. In optical platforms, linear optical elements are sufficient for implementing the entangling operations for the generation of cluster states as well as single-qubit-measurements. 

However, one major shortcoming that besets linear optical quantum computing is the fact that a {\it direct} Bell-state measurement (BSM), which is an entangling operation used to form a larger entangled state from smaller ones, is not deterministic~\cite{RP10,LCS99}. Adding to the shortcoming, photon loss is ubiquitous in all optical platforms and specifically in integrated optics~\cite{lossic}, which not only causes optical qubit states to leak out of the computational basis but also introduces dephasing or depolarizing noise into qubits, gate operations and readouts (measurements). Thus, the bare-bone measurement-based quantum-computing scheme in Ref.~\cite{RB01} is not tolerant against the aforementioned practical issues, and additional enhancements are necessary to ensure its successful execution in real experiments. To overcome indeterminism of BSM and quantum noise, we need fault-tolerant architectures that employ quantum error correction (QEC)~\cite{NC10, LD13}. With QEC, it is possible to achieve scalable linear optical quantum computing using lossy optical components provided the photon-loss level is below a certain threshold. This threshold value is dependent on the QEC codes and noise model considered in the fault-tolerant architecture~\cite{Steane03, Knill05, DHN06}.

Various kinds of linear-optical platforms have been proposed for quantum computing depending on the nature of variables used for encoding quantum information. Discrete-variable (DV) platforms manipulate level-structure properties of photons like polarization to encode quantum states. A BSM on DV qubits using linear optics is probabilistic with a success rate of $50\%$ without additional resources~\cite{LCS99, LC01}. Fault-tolerant architectures for linear optical quantum computing in Refs.~\cite{Niel04,DHN06, LDSB10, FT10,HFJR10, LHMB15} overcome indeterminism of entangling operations by the {\it repeat-until-success} strategy. However, if we wish to carry out, say, $m$ simultaneous and identical entangling operations successfully, the average resource overhead is $O(1/p_\mathrm{s}^m)$, which will result in an increase in overhead by a factor of $O(\Delta^m)$ that is exponential in $m$ when the success rate $p_\mathrm{s}$ of each entangling operation decreases by a factor of $\Delta$~\cite{LDSB10}. In our protocol, which we detail in Sec.~\ref{sec:mtqc}, the strategy with low success rates of entangling operations is used only at a certain stage unlike other mentioned architectures. This makes our protocol competitive in terms of resource overheads. Furthermore, because of such probabilistic entangling operations, these schemes would require {\it optical switches} to pass the successfully entangled states to the next step, which is known to contribute significant photon loss~\cite{BMOT15}. Alternatively, continuous-variable~(CV) platforms that employ coherent states described by a continuous parameter (amplitude)~\cite{JK02,RJM+03,LRH08,MR11} offers BSMs that can be nearly deterministic~\cite{JKL01}. Here, the success rate of a BSM grows with the coherent-state amplitude. The corresponding architectures for linear optical quantum computing require lower resource overheads, but are sensitive to photon loss and can only tolerate smaller thresholds~\cite{LRH08}. Also, there has been significant developments in using optical Gottesman-Kitaev-Preskill states for fault-tolerant quantum computing~\cite{FTO+18, Xanadu21}. These schemes need very large squeezing strengths ($>10$~dB) to implement gates with high precision.

Recently, there have been efforts to combine the DV and CV approaches for quantum computing~\cite{LJ13,OTJ19,OTJ20,Furu15,Take13}.
It was demonstrated that by using optical {\it hybrid} qubits, which are entangled states in the DV and CV domains, near-deterministic entangling operations can be implemented~\cite{LJ13,OTJ19}. Moreover, many shortcomings faced individually by CV and DV qubit-based schemes are overcome. Importantly, quantum computing with hybrid qubits reduces resource overheads and also improves the photon-loss tolerance~\cite{LJ13,OTJ19,OTJ20}.
By increasing the amplitude of the CV part, incurred resources can also be reduced. However, if the coherent amplitude of hybrid qubits is too large, the dephasing noise level in the presence of photon loss will also be commensurately too high~\cite{Phoenix90}. Reference~\cite{OTJ20} also supports the logic that quantum computing on a special cluster state known as Raussendorf--Harrington--Goyal~(RHG) lattice~\cite{RHG06,RHG07} built with only DV qubits could tolerate higher photon loss, albeit at higher resource costs. Larger cluster states like the RHG lattice is built by performing BSMs on smaller entangled states.

Besides probabilistic BSMs and photon loss, practical implementation of linear optical quantum computing is greatly hindered by indeterministic generation of multiphoton entangled states, such as the GHZ and cluster states. There are theoretical proposals for on-demand generations of such multiphoton entangled states~\cite{Wolf+05,Rudolph+09}. For instance, various multi-photon entangled states can be generated by coupling a multi-level ancillary system to a two-level photon-emitting source and performing sequential unitary operation on both the ancillary system and source~\cite{Wolf+05}. Here, the type of entangled states generated depends on the chosen unitary operation. In the experimental work~\cite{Besse20}, both the ancillary system and photon source are transmons, and are coupled \emph{via} a flux-tunable superconducting quantum interference device. Furthermore, by interacting the single-photon sources, such as quantum dots, with spin-1/2 states, the sources can be made to emit multi-photon entangled states~\cite{Rudolph+09}. Recent experimental realizations~\cite{SCS+16,TTF19} alternatively make use of  entanglement between the electron spin and polarization of photons emitted from optical excitations. An interesting point is that experimentally-realized three-photon and four-photon GHZ states respectively possess fidelities 0.9 and 0.82~\cite{Besse20}. We can also observe that as the number of photons in the GHZ state increases, the fidelity drops.

Motivated by the state-of-the-art techniques in deterministic generation of multiphoton entangled states, in this work we propose a \emph{multiphoton-qubit-based topological~quantum~computing} protocol~(MTQC) that uses multiphoton GHZ polarization states from deterministic sources to build RHG lattice. Furthermore, we demonstrate that our protocol provides an exceptionally high  tolerance against photon-loss that reaches $11.5 \%$. Considering GHZ states as the raw ingredients, we also show that the protocol is resource-efficient than all known \cite{HFJR10,DHN06,HHGR10,LPRJ15,Cho07,LRH08,LJ13} non-hybrid qubit-based schemes. We further encode each qubit of the RHG lattice with a multiqubit repetition code~\cite{NC10}---a concatenation of two QEC codes---to improve the photon-loss tolerance. Another salient and practically favorable feature of our protocol is that it requires only on-off detectors, unlike Ref.~\cite{LHMB15} that demands detectors that can resolve photon numbers and are thus more difficult to implement with competitive accuracy. Also photon number resolving nano-wire~\cite{Endo21} and transition-edge~\cite{Nehra19} based detectors cannot operate at room temperatures. However, on-off detectors can operate at room temperatures~\cite{CFLP+14}, which allows our MTQC protocol pave the way for scalable all-photonic~quantum~computing.

The rest of the article is organized as follows. In Sec.~\ref{sec:mtqc}, we describe our MTQC in detail. Next, in Sec.~\ref{sec:noise}, we discuss the noise model we consider throughout the work. Section~\ref{sec:qec} shall touch on the employment of concatenated QEC methods and their effects on further raising the threshold photon-loss rate that can be tolerated with our all-photonic architecture, where as the numerical simulation procedure of QEC is separately outlined in Appendix~\ref{sec:sim}. In Sec.~\ref{sec:result}, we present our results on the photon-loss thresholds, and the details concerning resource estimation, specifically the average number of three-qubit GHZ states consumed,  are provided in Sec.~\ref{sec:resource}. In Sec.~\ref{sec:compare}, we compare the results of our  MTQC with those of other linear optical quantum computing schemes. Finally, some pertinent discussion and conclusion are presented in Sec.~\ref{sec:conclusion}.

\section{Protocol}
\label{sec:mtqc}

The primary aim of MTQC is to build an RHG lattice for fault-tolerant quantum computing using multiphoton GHZ polarization states from deterministic sources and processing them with passive linear-optical elements like polarizing beam splitters, phase shifters, optical delay lines, optical switches and on-off detectors. To begin, multiphoton GHZ states are entangled by BSMs in an efficient manner to form two kinds of {\it resource states}. We point out that like other DV protocols, the repeat-until-success strategy with low-success-rate entangling operations is employed only at this stage to generate these resource states. After which, we perform a {\it collective  BSM}~\cite{LPRJ15} on multiphoton resource states, which is a near-deterministic entangling operation. 

The collective BSM (described in Sec.~\ref{subsec:nbsm}), is a crucial ingredient of our protocol. This is because the near-deterministic entangling operation requires only on-off detectors to boost its success rate of entangling arbitrarily close to 100\%. This is unlike some other BSMs~\cite{G11,EL14} that demand photon-number resolving detectors that can distinguish up to four photons to boost the success rate to 75\%, and the ability to resolve higher photon numbers is essential for further enhancements. Another salient feature of the collective BSM is that it needs no optical switching to entangle two optical qubits like the other DV protocols when using repeat-until-success strategy. These features make MTQC practically more attractive. It is also important to note that once we start using collective BSMs in our protocol, the repeat-until-success strategy, even if employed, shall not drastically increase resource overheads as the success rate of a collective~BSM is very high. This makes our protocol competitive in terms of resource overheads.

The RHG lattice built using BSMs with a boosted success rate of 75\% still cannot support fault-tolerant~quantum~computing as the failure rate must be lesser than 14.5\%~\cite{AAG+18}. To overcome this shortfall, there exists a proposal to purify the RHG lattice~\cite{HPDN18} by which the effective failure rate of BSMs can be brought down to 7\% from 25\%. While the purified RHG lattice can support fault-tolerant~quantum~computing, this approach has the disadvantage of reducing the effective size of the RHG lattice which will contribute to a large resource overhead. Also, in this situation the RHG lattice is less robust against dephasing errors. There is also an attempt to build the lattices with CV-based qubits in Ref.~\cite{MR11}. But this demands average photon numbers of CV qubits over 100 to build an RHG lattice that is of a sufficiently high fidelity for fault-tolerant computation. Such high average photon numbers are not achievable and lattices built under practical values~\cite{Cat21} (average photon number of~2) are far from fault-tolerant. Recently, there has been progress in the generation of  two-dimensional CV cluster-states without BSM~\cite{Larsen:2019deterministic,Asavanant:2019generation}. Another proposal~\cite{LHMB15} aims to build RHG lattices using BSMs with a boosted success rate of 75\% (or higher) by adding single photons or Bell pairs and employing photon number resolving detectors.  Here, the 14.5\% failure-rate mark is overcome by the repeat-until-success strategy to create entanglement between qubits similar to Ref.~\cite{DHN06}. 
Involvement of {\it tree clusters}~\cite{VBR06,VBR08} render the scheme fault-tolerant against BSM failure, but this comes at the cost of feed-forward measurements. Depending on the failure or success of BSM remaining qubits of tree cluster are measured in appropriate basis~\cite{LHMB15}. These feed-forward operations form the bottleneck in linear optical quantum computing. 

The state $\ket{\mathcal{C}_\mathcal{L}}$ of an RHG lattice is a unique 3D cluster state~\cite{RHG07} where qubits are entangled to their nearest neighbors represented by the edges of the RHG lattice. In general, a cluster state $\ket{\mathcal{C}}$ over a collection of qubits $\mathcal{C}$ is a state stabilized by the operators  $X_a\bigotimes_{b\in {\rm nh}(a)} Z_b$, where $a,b\in\mathcal{C}$, $Z_i$ and $X_i$ are the Pauli operators on the $i$th qubit, and nh($a$) denotes the adjacent neighborhood of qubit $a\in\mathcal{C}$:
\be
\ket{\mathcal{C}}=\prod_{\substack{a\in\mathcal{C}\\ a<b\in \rm{ nh}(a)}}\!\!\!\!\!\!\!\!\textrm{CZ}_{a,b}\,\bigotimes_{a'\in\mathcal{C}}\ket{+}_{a'}\,,
\label{eq:clus}
\ee
where $\ket{\pm}=(\ket{0}\pm\ket{1})/\sqrt{2}$ are eigenkets of $X$, while $\ket{0},\ket{1}$ are those of $Z$. The controlled-$Z$ unitary gate ${\rm CZ}_{a,b}$, which is an entangling operation, applies $Z$ on the target qubit $b$ if the source qubit $a$ is in the state $\ket{1}$. The successful action of ${\rm CZ}_{a,b}$ on the lattice qubits on sites $a$ and $b$ is represented by an edge in the lattice. The state $\ket{\mathcal{C}_\mathcal{L}}$ is currently the best available choice, to the best of our knowledge, to make linear-optical platform fault-tolerant; it can tolerate qubit loss~\cite{BS10,WF14}, probabilistic entangling operations~\cite{LDSB10,AAG+18}, dephasing and depolarizing noises~\cite{RHG06, RHG07,RH07}, all of which are peculiar to the platform. Furthermore, QEC and gate operations on $\ket{\mathcal{C}_\mathcal{L}}$ is topological in nature and thus offers the highest fault tolerance in the platforms where interactions  are restricted to those of the nearest neighbors~\cite{FG09}. 

In MTQC, the logical basis for an $l$-photon qubit is 
\be
\ket{0_l}\equiv\ket{\textsc{h}}^{\otimes l},~~\ket{1_l}\equiv\ket{\textsc{v}}^{\otimes l}.
\label{eq:raw}
\ee
where $\ket{\textsc{h}}$, $\ket{\textsc{v}}$ are the discrete orthonormal polarizations and are eigenkets of the $Z$ Pauli operator. An $r$-photon GHZ ket has the form $\ket{{\rm GHZ}_r}\propto\ket{\textsc{h}}^{\otimes r}+\ket{\textsc{v}}^{\otimes r}$ (up to normalization). 
Once there is a continuous and reliable supply of $\ket{{\rm GHZ}_r}$ from deterministic sources, the following prescription forms the stages of our protocol to create $\ket{\mathcal{C}_\mathcal{L}}$ and perform topological fault-tolerant~quantum~computing on it:
\begin{enumerate}
\item The foremost step is to create multiphoton three-qubit resource states using $\ket{{\rm GHZ}_r}$ and BSMs.
\item Near-deterministic collective BSM  are performed on these resource states to form {\it star cluster states}.
\item The star clusters then undergo collective BSM to form layers of  $\ket{\mathcal{C}_\mathcal{L}}$.
\item Finally, the qubits are measured layer by layer in a suitable basis to effect both QEC and Clifford-gate operations on the logical states of $\ket{\mathcal{C}_\mathcal{L}}$. Initialization of  $\ket{\mathcal{C}_\mathcal{L}}$ to certain logical states and magic-state distillation are also possible by measurements which complete the universal set of gates for quantum computing.
\end{enumerate}

Certain aspects of MTQC do bear some resemblance with the protocol in Ref.~\cite{OTJ19}, which uses hybrid qubits as basic ingredients. However, this resemblance is only superficial. Since we use GHZ states as basic ingredients in MTQC, the very process of generating three-qubit resource states is different. The kind of entangling operations and apparatus employed are also very different. Here, we need on-off detectors whereas the protocol in Ref.~\cite{OTJ19} requires photon-number-parity detectors. Additionally, MTQC exploits the concatenation of error-correcting codes to improve the tolerance against photon loss. In what follows, all stages of the MTQC scheme are discussed in detail.

\subsection{Collective Bell-state measurement on multiphoton qubits}
\label{subsec:nbsm}
The Bell states of multiphoton qubits, each with $n$ modes, are $\ket{\phi^\pm_n}=(\ket{0_n0_n}\pm\ket{1_n1_n})/\sqrt{2}$, $\ket{\psi^\pm_n}=(\ket{0_n1_n}\pm\ket{1_n0_n})/\sqrt{2}$. Interestingly, Bell states of multiphoton qubit can be decomposed as Bell states of individual photon modes, that is 
$\ket{\phi^\pm}=(\ket{\textsc{hh}}\pm\ket{\textsc{vv}})/\sqrt{2}$, $\ket{\psi^\pm}=(\ket{\textsc{hv}}\pm\ket{\textsc{vh}})/\sqrt{2}$ as follows~\cite{LPRJ15}:
\bea
\ket{\phi^+_n}&=&\frac{1}{\sqrt{2^{n-1}}}\sum_j\mathcal{P}_n\left[\ket{\phi^+}^{\otimes n-2j}, \ket{\phi^-}^{\otimes 2j}\right] \nonumber\\
\ket{\phi^-_n}&=&\frac{1}{\sqrt{2^{n-1}}}\sum_j\mathcal{P}_n\left[\ket{\phi^+}^{\otimes n-2j-1}, \ket{\phi^-}^{\otimes 2j+1}\right] \nonumber\\
\ket{\psi^+_n}&=&\frac{1}{\sqrt{2^{n-1}}}\sum_j\mathcal{P}_n\left[\ket{\psi^+}^{\otimes n-2j}, \ket{\psi^-}^{\otimes 2j}\right] \nonumber\\
\ket{\psi^-_n}&=&\frac{1}{\sqrt{2^{n-1}}}\sum_j\mathcal{P}_n\left[\ket{\psi^+}^{\otimes n-2j-1}, \ket{\psi^-}^{\otimes 2j+1}\right] 
\eea
Here, for two states (or operators) $A$ and $B$ that are $i$- and $(n-i)$-partite respectively ($i \leq n$) and invariant under permutations of their supports, we define $\mathcal{P}_n[A, B] := \sum_{I \in S_{n,i}} A_I \otimes B_{\mathbb{Z}_n \setminus I}$, where $\mathbb{Z}_n := \left\{1, \cdots, n\right\}$, $S_{n,i} := \left\{I \subseteq \mathbb{Z}_n : |I| = i \right\}$, and $A_I$ indicates the state (or operator) $A$ supported on $I$.
As an example, if $A$ and $B$ are single-partite operators, $\mathcal{P}_4[A^{\otimes 2}, B^{\otimes 2}]=A\otimes A\otimes B\otimes B + A\otimes B\otimes A\otimes B + A\otimes B\otimes B\otimes A + B \otimes A \otimes A \otimes B + B\otimes A\otimes B\otimes A + B\otimes B\otimes A\otimes A$.

\begin{figure}[t]
	\includegraphics[width=.9\columnwidth]{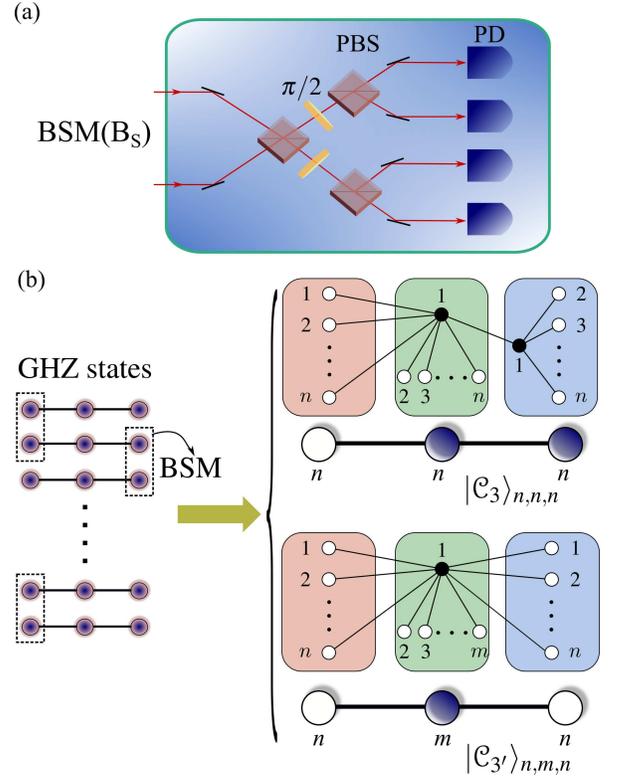}
	\caption{(a) Bell-state measurement setup, ${\rm B_S}$ consists of three polarization beam splitters (PBS), two $\pi/2$-rotators and four on-off photon detectors (PD). ${\rm B_S}$ is deemed to be successful when one of the first two PDs and one of the last two PDs click simultaneously. When successful, the setup can discriminate only two Bell states from four of them.  Therefore, the success rate of ${\rm B_S}$ is $1/2$. (b) Resource states $\ket{\mathcal{C}_{3}}_{n,n,n}$ and $\ket{\mathcal{C}_{3^\prime}}_{n,m,n}$ [see Eq.~\eqref{eq:3clus}] can be generated using GHZ states from deterministic sources. We depict $\ket{{\rm GHZ}_3}$, but  $\ket{{\rm GHZ}_r}$, where $r>3$ can be used in principle. While all qubits of  $\ket{\mathcal{C}_{3}}_{n,n,n}$ have $n$ photons, the first, second and third qubits of  $\ket{\mathcal{C}_{3^\prime}}_{n,m,n}$ have $n$, $m$ and $n$ photons, respectively. The unfilled circles refer to the qubits on which the Hadamard operation is carried out as explained below Eq.~\eqref{eq:3clus}, and a solid line represents the existence of entanglement between the qubits. More refined pictures concerning the respective formations of $\ket{\mathcal{C}_{3}}_{n,n,n}$ and $\ket{\mathcal{C}_{3'}}_{n,m,n}$ states (both unitarily equivalent to graph states) are supplied here, where we intuitively note that $\ket{\mathcal{C}_{3}}_{n,n,n}$ is a result of a controlled-$Z$ operation of two kinds of multiphoton GHZ states and $\ket{\mathcal{C}_{3'}}_{n,m,n}$ is itself a large multiphoton GHZ state. While a general formula for the number of GHZ states consumed to generate  $\ket{\mathcal{C}_{3}}_{n,n,n}$ and $\ket{\mathcal{C}_{3^\prime}}_{n,m,n}$ is absent, this may be explicitly calculated from the iterative procedure graphically presented in Fig.~\ref{fig:c3} and in Appendix~\ref{app:GHZ_r}. Also see Sec.~\ref{sec:resource} for examples. 
	}
	\label{fig:bs-3qubit}
\end{figure}

From the above expression it is clear that if the states  $\ket{\phi^-}$ and $\ket{\psi^-}$ can be distinguished at mode level, all the four Bell states can be identified at the logical level with success rate $1-2^{-n}$. Thus, as we add more photon-modes to qubits, the success rate of BSMs on them improves and becomes near-deterministic. 
The states $\ket{\phi^-}$ and $\ket{\psi^-}$ can be unambiguously distinguished by the BSM setup $B_S$ shown in Fig.~\ref{fig:bs-3qubit}. If  $B_S$ registers even (odd) number of $\ket{\phi^-}$'s, the corresponding state at the logical level is $\ket{\phi^+_n}$ ($\ket{\phi^-_n}$). On the other hand if even (odd) number of $\ket{\psi^-}$'s are registered, the  corresponding state at the logical level is $\ket{\psi^+_n}$ ($\ket{\psi^-_n}$).

This scheme is robust against failure of bare BSM due to photon loss.  
Employing this scheme of near-deterministic collective BSM has another advantage over protocols which use multiple attempts of BSMs to entangle two optical smaller cluster states. In the latter case, the bare BSMs must be applied in a sequence and thus there is associated waiting time for each of them during which photons may be lost. Also, when one of the BSMs succeed, the left over photons must be trimmed from the star cluster state. Thus, the process also demands extensive usage of delay lines and switching networks. But in our case all the BSMs are applied simultaneously removing the necessity for delay lines and switching networks, and the associated noise.

\begin{figure*}[t]
	\includegraphics[width=2\columnwidth]{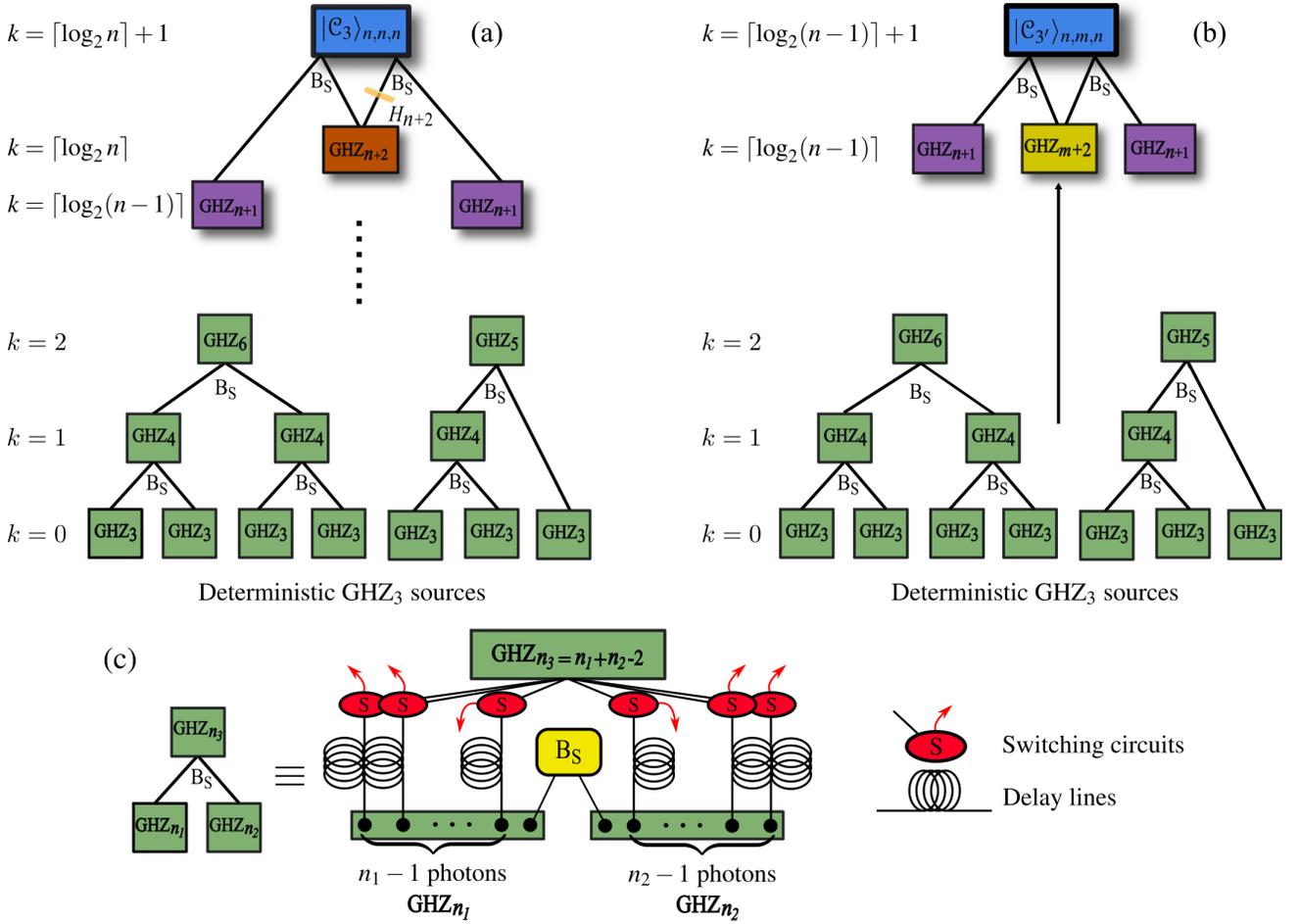}
	\caption{Generation of resource states for MTQC. (a) The resource state $\ket{\mathcal{C}_{3}}_{n,n,n}$ (see Eq.~(\ref{eq:3clus})) can be probabilistically generated by performing BSMs on  $\ket{{\rm GHZ}_3}$'s from deterministic sources in $\lceil\log_2 n\rceil+1$ steps. It takes  $k=\lceil\log_2 n\rceil$ steps to generate $\ket{{\rm GHZ}_{n+2}}$'s. In the $(k+1)$th step two $\ket{{\rm GHZ}_{n+1}}$'s are added from the  step corresponding to suitable value of $k$.  In this step, Hadamard unitary operations on the last photon of the $\ket{{\rm GHZ}_{n+2}}$ ($H_{n+2}$) is performed before feeding it to ${\rm B_S}$. This step involves two ${\rm B_S}$'s and $\ket{\mathcal{C}_{3}}_{n,n,n}$ is formed only when both are successful. (b) To create a $\ket{\mathcal{C}_{3^\prime}}_{n,m,n}$ two $\ket{{\rm GHZ}_{n+1}}$'s and a $\ket{{\rm GHZ}_{m+2}}$ (from step of suitable $k$) are entangled as shown with out any Hadamard operation. Upon both ${\rm B_S}$'s being successful, $\ket{\mathcal{C}_{3^\prime}}_{n,m,n}$ is created. At the $k$th step, the largest possible GHZ state has $2^k+2$ photons and other smaller GHZ states can be generated by entangling GHZ states from whichever previous steps, as depicted at $k=2$.  Smaller possible states are listed in Tab.~\ref{tab:ghz}. Although the flowchart starts with $\ket{\rm GHZ}_3$ as the basic ingredients, $\ket{{\rm GHZ}_r}$ of any $r>3$ may also be used in principle. In doing so, the value of $k$ required to generate resource states can be reduced. In the case of encoding the qubits of $\ket{\mathcal{C}_\mathcal{L}}$ with three-qubit repetition QEC codes, the $\ket{{\rm GHZ}_{m+2}}$'s in the $k$th step is replaced by the $\ket{{\rm enc}}$'s (see Eq.~(\ref{eq:enc})) to form  $\ket{\mathcal{C}_{3^\prime}}_{{\rm enc}}$'s in~Eq.~\eqref{eq:c3enc}. Refer to Fig.~\ref{fig:enc} for the process of creating $\ket{{\rm enc}}$. In the $(k+1)$th step $\ket{{\rm GHZ}_{n+2}}$ is added from the step(s) corresponding to lower value(s) of $k$.  (c) Each action of ${\rm B_S}$ involves delay lines and switching networks. As an example, consider $\ket{{\rm GHZ}_{n_1}}$ and  $\ket{{\rm GHZ}_{n_2}}$ being entangled to form  $\ket{{\rm GHZ}_{n_3=n_1+n_2-2}}$. The delay lines are employed to slow the passage of $n_3=n_1+n_2-2$ photons until the action of the ${\rm B_S}$ is complete. The switching network routes the leftover $n_3$ photons to the next level if ${\rm B_S}$ is successful. Otherwise, the photons are discarded. }
	\label{fig:c3}
\end{figure*}

\subsection{Resource states}
Our protocol for MTQC begins with the creation of the following two kinds of multiphoton resource states,
\bea
\label{eq:3clus}
\ket{\mathcal{C}_3}_{n,n,n}&=&
\frac{1}{2}\big(\ket{0_n0_n0_n}+\ket{0_n0_n1_n}+\ket{1_n1_n0_n}-\ket{1_n1_n1_n}\big)\,,\nonumber\\
\ket{\mathcal{C}_{3^\prime}}_{n,m,n}&=&
\frac{1}{\sqrt{2}}\big(\ket{0_n0_m0_n}+\ket{1_n1_m1_n}\big),
\eea
which are three-qubit entangled states at the logical level.  Throughout the article, we suppose that $\ket{\mathcal{C}_{3}}_{n,n,n}$ has $n$ polarization photons in all qubits where as $\ket{\mathcal{C}_{3^\prime}}_{n,m,n}$ has  $n$, $m$, and $n$ polarization photons in the first, second and third qubits, respectively.
One can verify that $\ket{\mathcal{C}_3}$ is the result of a Hadamard operation on the first qubit of the three-qubit linear cluster state ${\rm CZ}_{1,2}{\rm CZ}_{2,3}\ket{+_n+_n+_n}$. On the other hand, $\ket{\mathcal{C}_{3^\prime}}$ is obtained by a Hadamard operation on the first and third qubits of the same three-qubit linear cluster state and is a logical GHZ state.

The resource states are created by entangling 
$r$-photon GHZ states from deterministic sources using ${\rm B_S}$, as shown in Fig.~\ref{fig:bs-3qubit}. This way of creating resource states using  ${\rm B_S}$'s is possible when $r\geq3$. Due to the usage of ${\rm B_S}$'s, the process is probabilistic and the required states are generated only when all the ${\rm B_S}$'s succeed. If one of the ${\rm B_S}$'s fails, the state is discarded and the process is restarted; that is, the repeat-until-success strategy is employed. This is also reflected in the resource overhead calculations carried out in Sec.~\ref{sec:resource}. A successful ${\rm B_S}$ can distinguish $\ket{\phi^+}$ and $\ket{\phi^-}$. If the outcome is $\ket{\phi^-}$, a feed-forward $Z$ operation would be necessary on a photon in resultant GHZ state. However, there is no need for physical implementation of this $Z$ operation as such a feed-forward procedure can be realized by updating the {\it Pauli frame}~\cite{DHN06}. Accordingly, measurement results on the photons must be interpreted in corroboration with the Pauli frame. On the other hand, failed BSMs make the involved GHZ states mixed, thus they cannot be restored by feed-forward operations without additional resources. One solution is to use a suitable error correction scheme to handle the failures, the discussion of which is beyond the scope of this article.

In this work, we consider $r=3$, the smallest possible value, considering its experimental availability using current technology~\cite{Besse20}. In principle, $\ket{{\rm GHZ}_r}$'s with $r>3$ can be used at this stage, which would not only reduce the required number of ${\rm B_S}$, but also improve resource efficiency (average number of incurred $\ket{{\rm GHZ}_r}$'s) for generating $\ket{\mathcal{C}_3}_{n,n,n}$ and $\ket{\mathcal{C}_{3^\prime}}_{n,m,n}$. However, a larger $r$ would imply a poorer average fidelity of the GHZ states generated experimentally~\cite{Besse20}.  
The state $\ket{\mathcal{C}_3}_{n,n,n}$ is created by entangling two  $\ket{{\rm GHZ}_{n+1}}$'s and $H_{n+2}\ket{{\rm GHZ}_{n+2}}$, where  $H_{j}$ is the Hadamard operator acting on the $j$th photon, using ${\rm B_S}$'s as shown in the final step of Fig.~\ref{fig:c3}(a). The Hadamard operation is achieved by passing the polarization photons through a $\pi/2$-rotator. Similarly, procedure to generate $\ket{\mathcal{C}_{3^\prime}}_{n,m,n}$ is shown in the Fig.~\ref{fig:c3}(b). In this case no Hadamard operation is involved. 

The states $\ket{{\rm GHZ}_{n+1}}$ can be efficiently created by acting ${\rm B_S}$'s on $\ket{{\rm GHZ}_3}$'s in $\lceil\log_2(n-1)\rceil$ steps as illustrated in Fig.~\ref{fig:c3}. Here, $\lceil.\rceil$ represents the ceiling of a number. For example, when the step number $k=3$, up to $\ket{{\rm GHZ_{10}}}$ (where $n+1=10$) can be created (other possible states are listed in Tab.~\ref{tab:ghz}). The resource states $\ket{\mathcal{C}_3}_{n,m,n}$ and $\ket{\mathcal{C}_{3^\prime}}_{n,m,n}$ are created with success rates $(2^{\lceil\log_2n\rceil+2})^{-1}$ and $(2^{\lceil\log_2(n-1)\rceil+2})^{-1}$, respectively. At this stage of the protocol, delay lines and optical switches are employed. Delay lines are essential to delay the passage of photons that are not undergoing measurement until the action of the current ${\rm B_S}$ is complete. Therefore, quicker ${\rm B_S}$'s permit shorter delay lines. Optical switches are needed to control the flow of photons through these delay lines, choose the successful outputs of the ${\rm B_S}$'s, and send the larger GHZ states to the next step. Hence, many generation steps would entail a large number of optical switches and longer delay lines. Therefore, our resource-state generation protocol aims to minimize the usage of delay lines and optical switches that significantly contribute to photon loss. On the other hand, when deterministic sources capable of producing high-fidelity $\ket{{\rm GHZ}_r}$'s with $r\geq 4$ are used, resource states can be generated with a smaller number of time steps. The total number of steps to complete the generation of $\ket{\mathcal{C}_3}_{n,n,n}$ $\left(\ket{\mathcal{C}_{3'}}_{n,m,n}\right)$ is $\lceil\log_2 n\rceil+1$ $\left(\lceil\log_2(n-1)\rceil+1\right)$, when $m+1<n$.

Let us consider the creation of $\ket{\mathcal{C}_3}_{8,8,8}$ and $\ket{\mathcal{C}_{3^\prime}}_{8,2,8}$ as an example. To create this state, we need to entangle $\ket{{\rm GHZ}_{9}}$, $H_{10}\ket{{\rm GHZ}_{10}}$ and $\ket{{\rm GHZ}_{9}}$. All the required GHZ states are generated at ($k=3$)th step using $\ket{{\rm GHZ}_{3}}$'s. Upon having successful ${\rm B_S}$ on the 9th photon of $\ket{{\rm GHZ}_{9}}$ and 1st photon of $H_{10}\ket{{\rm GHZ}_{10}}$, and on the 10th photon of $H_{10}\ket{{\rm GHZ}_{10}}$ and 1st photon of $\ket{{\rm GHZ}_{9}}$, we create $\ket{\mathcal{C}_3}_{8,8,8}$ in $(k=4)$th step. Similarly, when no $H_{10}$ is performed and replacing  $\ket{{\rm GHZ}_{10}}$ with $\ket{{\rm GHZ}_{4}}$ we end up having $\ket{\mathcal{C}_{3'}}_{8,2,8}$.

\begin{figure}[t]
\includegraphics[width=0.9\columnwidth]{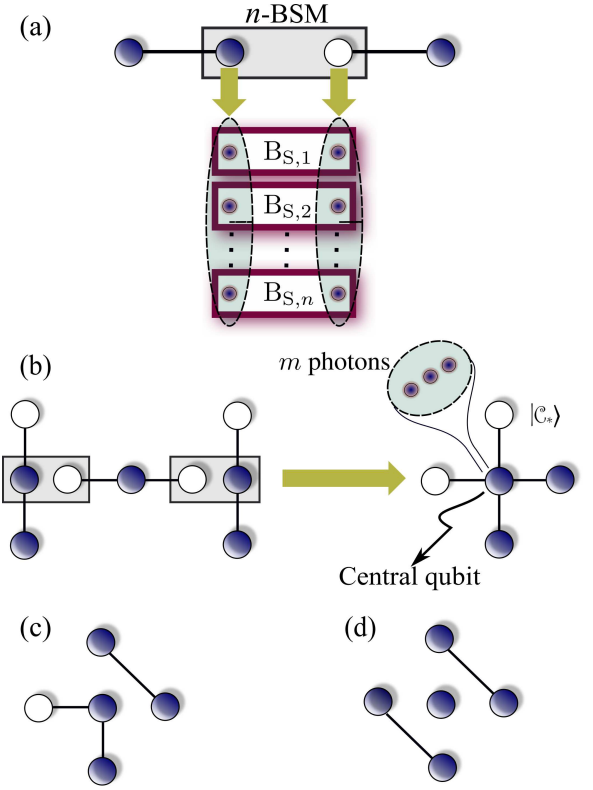}
\caption{(a) An $n$-BSM is used to create entanglement between two resource states. It consists of a cascade of $n$ ${\rm B_S}$'s acting on photons from two different $n$-photon qubits as shown. The success rate of $n$-BSM increases with $n$ as $1-2^{-n}$ (refer to Sec.~\ref{subsec:nbsm}), which makes it a near-deterministic entangling operation. (b)~The star cluster state $\ket{\mathcal{C}_\ast}$ is generated by performing two $n$-BSM on qubits of two $\ket{\mathcal{C}_3}_{n,n,n}$ and a $\ket{\mathcal{C}_{3^\prime}}_{n,m,n}$ as shown. The desired $\ket{\mathcal{C}_\ast}$ is formed only when both $n$-BSMs are successful. (c) The resultant state when one of $n$-BSMs fails and (d) that when both fail. The solid line represents the existence of entanglement between the qubits. 
The surrounding qubits are consumed by $n$-BSMs for creating entanglement between the central qubits (containing $m$ photons) in the formation of  $\ket{\mathcal{C}_\mathcal{L}}$. All central qubits contain lesser photons so that photon-loss induced dephasing is reduced (see Sec.~\ref{sec:noise}). }
\label{fig:nBSM}
\end{figure}

\subsection{Star cluster states}
\label{sec:star}
After explaining the procedure to create resource states in detail, we shall now discuss the formation of the star cluster state $\ket{\mathcal{C}_\ast}$. Here, we use near-deterministic  $n$-Bell state measurement ($n$-BSM)~\cite{LPRJ15} which is a cascade of $n$ ${\rm B_S}$'s, as shown in Fig.~\ref{fig:nBSM}(a), to entangle the resource states. An $n$-BSM fails when all the constituent ${\rm B_S}$'s fail. Therefore, the success rate of $n$-BSM is $1-1/2^n$ and arbitrarily approaches 1 with increasing $n$. The working principle of $n$-BSM is explained in Sec.~\ref{subsec:nbsm}. Two $\ket{\mathcal{C}_3}_{n,n,n}$'s and a $\ket{\mathcal{C}_{3^\prime}}_{n,m,n}$ are entangled using two $n$-BSMs to form a $|\mathcal{C}_\ast\rangle$ as shown in the Fig.~\ref{fig:nBSM}(b). A $|\mathcal{C}_\ast\rangle$ is formed only when both the $n$-BSMs are successful in the process. In other failed cases, the desired $|\mathcal{C}_\ast\rangle$ is not formed and the resulting states are distorted as shown in Fig.~\ref{fig:nBSM}(c) and Fig.~\ref{fig:nBSM}(d). A successfully generated $|\mathcal{C}_\ast\rangle$ shall have $m$ photons in the central qubit and $n$ photons in the surrounding qubits. Having a larger $n$ is desirable as the success rate of the $n$-BSM improves when $|\mathcal{C}_\ast\rangle$'s are entangled to form layers of $|\mathcal{C}_\mathcal{L}\rangle$. While having $m>1$ suppresses the probability of qubit loss on $\ket{\mathcal{C}_\mathcal{L}}$ (as qubit is redundantly encoded with multiple photons) during photon loss, it also invites stronger dephasing on the qubit  as described in Sec.~\ref{sec:noise}. 

\begin{figure}[t!]
	\includegraphics[width=1\columnwidth]{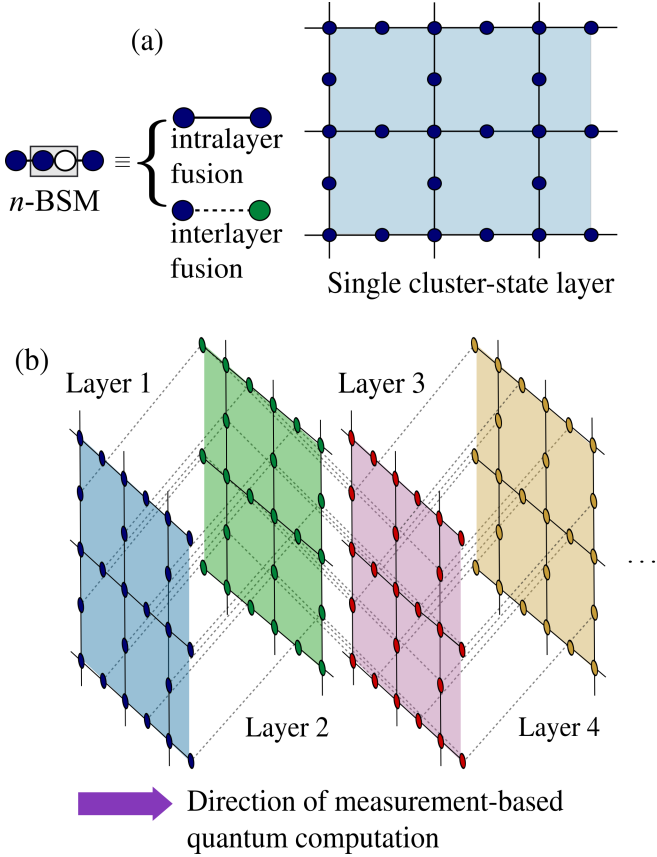}
	\caption{(a)~Each layer of RHG lattice is generated  by entangling the star cluster states~$\ket{\mathcal{C}_*}$ using $n$-BSMs. (b)~In implementation schemes where multiple layers that form $\ket{\mathcal{C}_\mathcal{L}}$ can be created in time through interlayer $n$-BSMs.}
	\label{fig:layers}
\end{figure}

The $n$-BSM operations employed in our protocol scheme are surely prone to failures.~At this point we introduce two subvariants of MTQC. In the first subvariant, referred to as MTQC-1, we do not employ any optical switches and use whatever star cluster states upon failure of $n$-BSMs. In the second subvariant, namely MTQC-2, we use optical switches circuit to choose intact $|\mathcal{C}_\ast\rangle$'s and discard the distorted ones. This will make the photons to pass through  another delay line and switch. Thus, MTQC-1 adds one extra time step in creating $|\mathcal{C}_\ast\rangle$ while MTQC-2 adds two. To be explicit, until now the process of creating $\ket{\mathcal{C}_\mathcal{L}}$ has taken place in $k+1$ and $k+2$ time steps with MTQC-1 and MTQC-2, respectively. We shall study and compare the performances of both these subvariants in terms of photon-loss tolerance and resource efficiency.

\subsection{Layers of RHG lattice}
The $|\mathcal{C}_\ast\rangle$'s  are entangled using $n$-BSMs to form layers of  $\ket{\mathcal{C}_\mathcal{L}}$ as shown in Fig.~\ref{fig:layers}.  During the process, surrounding qubits of each $|\mathcal{C}_\ast\rangle$ are consumed by the $n$-BSMs and the central qubit stays on the lattice. Upon a successful $n$-BSM, edges (representing entanglement) between the central qubits are formed. 
In MTQC-2 intact $|\mathcal{C}_\ast\rangle$'s are generated.
If an $n$-BSM fails, the corresponding edge between the central qubits will be missing and such situations are handled by QEC on $\ket{\mathcal{C}_\mathcal{L}}$. For details on how missing edges on $\ket{\mathcal{C}_\mathcal{L}}$ are handled, we refer the Reader to Refs.~~\cite{LDSB10,AAG+18}. However, in MTQC-1, distorted $\ket{\mathcal{C}_\ast}$'s are allowed and this gives rise to  $\ket{\mathcal{C}_\mathcal{L}}$ with {\it diagonal} edges (refer also to Fig.~2(c) of Ref.~\cite{OTJ19}) along with missing edges. The diagonal edges distort the local RHG structure, which in turn makes usual error syndrome extraction impossible. This too, like the problem of missing edges, is overcome by QEC where qubits associated with diagonal edges are removed. Therefore, the state ket $\ket{\mathcal{C}_\mathcal{L}}$ generated in MTQC-1 would have more missing qubits (explained in detail in Sec.~\ref{sec:result}), and is thus of a poorer quality. We emphasize that at this stage of formation of layers in MTQC, no optical switch  is involved in both the protocol subvariants. 

In MTQC-1 the formation of $|\mathcal{C}_\mathcal{L}\rangle$ and $|\mathcal{C}_\ast\rangle$ takes place simultaneously; that is, both happen in the $(\lceil\log_2 n\rceil+2)$th time step. But in MTQC-2 the formation of $|\mathcal{C}_\mathcal{L}\rangle$ happens in the next time step after the intact $|\mathcal{C}_\ast\rangle$ are formed. Therefore, $|\mathcal{C}_\mathcal{L}\rangle$ is formed in $(\lceil\log_2 n\rceil+3)$th time step. Another point to be noted is that the qubits of $|\mathcal{C}_\ast\rangle$ that take part in entangling to future layer must wait for an extra time step. This completes the protocol for generating $\ket{\mathcal{C}_\mathcal{L}}$. 

\subsection{Universal quantum computing on RHC lattice}
Following the RHG-lattice generation, where a faulty $\ket{\mathcal{C}_\mathcal{L}}$ with missing edges and phase-flip errors is created, topological fault-tolerant~quantum~computing is carried out by making sequential single-qubit measurements in $X$ and $Z$ bases as dictated by the quantum gates being implemented. Measurements in the $Z$ basis remove qubits from  $\ket{\mathcal{C}_\mathcal{L}}$, creating {\it defects} which also creates logical states of lattice. These defects are braided to achieve two-qubit logical operations topologically~\cite{RHG06,RHG07}. The $X$-basis measurement outcomes  provide error syndromes and also effect Clifford gates on the logical states. The universal set of operations for quantum computing is complete with the inclusion of {\it magic-state distillation} for which measurements on the chosen qubits are carried out in the $(X\pm Y)/\sqrt{2}$ basis~\cite{RHG06,RHG07}. Practically, $Z$-basis measurements on qubits are possible by measuring the polarization of any photon belonging to the qubit in the $z$ direction. On the other hand, an $X$-basis measurement outcome of a lattice-qubit is given by the parity of $X$-basis measurements of the constituent photons. 

{\it Logical errors} on $\ket{\mathcal{C}_\mathcal{L}}$ occur when a chain of $Z$ operators connects two defects or encircles a defect. Code distance, $d$~(lattice size) is the minimum number of $Z$ operations on the lattice qubits such that two defects are connected. In the following we explain how random $Z$ operators (errors) on optical qubits  happen due to photon-loss. These errors coupled with wrong inference in decoding during QEC go undetected and shall lead to logical errors~\cite{RHG06,RHG07}. Also refer to Appendix~\ref{sec:sim} for more details on logical errors. 
The logical errors are faulty gate operations and can be minimized by choosing sufficiently large $d$. If $p_{\rm L}$ denotes the logical error rate, our aim is to reduce it to a target error rate, $p_{\rm L}^{\rm targ}$  set by the end user.

\section{Noise model}
\label{sec:noise}
Apart from photon loss being a major source of errors~\cite{RP10}, the lattice $\ket{\mathcal{C}_\mathcal{L}}$ built with linear optics suffer from missing edges due to probabilistic entangling operations. In this section, we study the effect of photon loss on multiphoton qubits, success rate of entangling operation and consequently on the MTQC protocol. Suppose that the overall photon-loss rate suffered by each photon due to imperfect optical components is $\eta$ and the initial state of an $l$-photon qubit is defined by $\ket{\Psi^l}=\alpha\ket{\textsc{h}}^{\otimes l}+\beta\ket{\textsc{v}}^{\otimes l}$. When $\eta$ is nonzero, the state of the multiphoton qubit is~\cite{LPRJ15}
\begin{align}
	\widetilde{\rho}_{l}=&\,\ket{\Psi^l}(1-\eta)^l\bra{\Psi^l}+\frac{1}{2}\sum^l_{q=1}\eta^q(1-\eta)^{l-q}\nonumber\\
	&\,\times\mathcal{P}_l[\ket{\Psi^{l-q}}\bra{\Psi^{l-q}}+\ket{\Psi^{l-q}_-}\bra{\Psi^{l-q}_-},\ket{\textsc{vac}_q}\bra{\textsc{vac}_q}]\,,
\end{align}
where $\ket{\Psi^{l-q}_-}=\alpha\ket{\textsc{h}}^{\otimes l-q}-\beta\ket{\textsc{v}}^{\otimes l-q}$ and $\ket{\textsc{vac}_q}$ is a vacuum state. The probability of an $l$-photon qubit losing a photon or more is $1-(1-\eta)^l$. When photons are lost, with probability 0.5 the state possesses the component $\ket{\Psi^l_-}\bra{\Psi^l_-}$, that is, undergoes dephasing.

If $0\leq\eta\leq1$ is the overall photon-loss rate due to imperfect GHZ source ($\eta^{\rm soc}$), delay lines ($\eta^{\rm dly}$), optical switching network~($\eta^{\rm swc}$) and detectors ($\eta^{\rm det}$), the rate of dephasing on the multiphoton lattice qubits due to photon loss is
\bea
\label{eq:pz}
p_Z=\frac{1-(1-\eta)^{l}}{2}.
\eea
If photon losses in different components of quantum computing are statistically independent, we have the relation 
\begin{equation}
\eta=1-(1-\eta^{\rm soc})(1-\eta^{\rm dly})(1-\eta^{\rm swc})(1-\eta^{\rm det})
\label{eq:comp_eta}
\end{equation}
that associate the overall photon-loss rate to the relevant individual component loss rates. For the delay lines $(1-\eta^{\rm dly})= \E{-c\,\tau_0\,\kappa/L_0}$, where $L_0=22$~km is attenuation length of optical fiber for the standard telecom wavelength of 1550~nm~\cite{Zwerger_2017}, $c=2\times10^5$~km/s is the speed of light in, say, acrylic (PMMA) optical fibers, $\tau_0=150\times10^{-9}$~s~\cite{RP10} is the time duration to complete one BSM, which can be made smaller using electro-optical modulators, and $\kappa$ is the total time steps (in units of $\tau_0$) a photon of lattice-qubit; that is, central qubit of $\ket{\mathcal{C}_\ast}$ spends before being measured during fault tolerant quantum computing. Also, each photon has to pass through a network of $\kappa$ switches. Therefore, $\eta^{\rm swc}=1-(1-\eta^{\rm s})^{\kappa}$, where $\eta^{\rm s}$ is photon-loss rate through one optical switch.
It is important to note from Eq.~\eqref{eq:pz} that photon-loss introduces larger rate of dephasing on qubits with larger number of photons. 

The ${\rm B_S}$ operates on the lossy states and when input photons are lost the failure rate increases to $1-(1-\eta)^2/2$. An $n$-BSM would fail only when all the constituent ${\rm B_S}$'s fail. Therefore, the failure rate of an $n$-BSM due to photon loss is
\begin{equation}
	p_\mathrm{f}=\left[1-\dfrac{(1-\eta)^2}{2}\right]^n\underbrace{\cong}_{\eta\ll1}\left(\dfrac{1}{2}+\eta\right)^n\,.
	\label{eq:nBSM_fail}
\end{equation}
From the above expression, we observe that when $\eta$ is $O(10^{-2})$ and $n$ is large, $p_{\rm f}$ is not very sensitive to photon loss.

We point out that like other DV optical schemes \cite{HFJR10}, photon loss does not necessarily imply lattice-qubit loss. The probability that photon loss leads to lattice-qubit loss, $\eta^m$, is much smaller than $p_{\rm f}$ for $\eta\sim10^{-2}$, $n\geq5$ and $m\geq2$ considered in this work. Therefore, $n$-BSM failure has a dominant effect over qubit loss when calculating {\it logical error rate} on $\ket{\mathcal{C}}_\mathcal{L}$ and the latter can be neglected. However, having large $m$ is not favorable as it invites stronger dephasing as inferred from Eq.~(\ref{eq:pz}). To mitigate this issue we set $m=2$ through out the work, which also sufficient to neglect the effect of qubit loss.

The MTQC is operated at the same photon-loss rate $\eta$ on all qubits and the number of photons in lattice qubits is $m=2$. In this case $\kappa<k$ and it is crucial to note that the lattice qubits always spend  $\kappa=3~(4)$ time steps in MTQC-1~(MTQC-2) before being measured. Now one can appreciate that our protocol to generate $\ket{\mathcal{C}_{3^\prime}}_{n,2,n}$ makes sure that the photons of lattice qubits passes through least number of optical components.  Therefore, for a fixed $\eta$ component-wise photon-loss tolerence of MTQC is enhanced. Also, note that before performing $n$-BSMs to form $\ket{\mathcal{C}_\mathcal{L}}$, the \emph{surrounding} qubits of a $|\mathcal{C}_\ast\rangle$ has to pass through a larger number of lossy components as $n>m$ and therefore have a larger photon-loss rate. Moreover, qubits that are part of the $n$-BSM between the current and future layers spend an extra time step and thus suffer from stronger losses.
However, this time delay depends on the physical architecture that runs measurement-based quantum computation along with other technological details; in principle, different layers can be generated simultaneously. Therefore, we hereby neglect the influence of time delays on $\eta$, which is reasonable if the time delay is sufficiently shorter than $-\left(L_0/c\right) \ln\eta$.

For the star cluster states, qubit dephasing owing to photon loss happens \emph{locally} on the central and other qubits. 
In addition, the surrounding qubits are consumed during an $n$-BSM and noise owing to photon loss can be dealt with by suitably encoding these qubits. Investigation of this procedure is a subject matter of our future work that is beyond the scope of this article. In this work, we consider only the dephasing noise due to photon loss on the central qubits. One can also consider other kinds of amplitude-damping noise~\cite{OSB13,OSB15t,SB08,Srik18} to evaluate the performance our MTQC protocols.

\section{Encoding lattice qubits with the repetition code}
\label{sec:qec}

As described in the previous section, photon-loss leads to dephasing on the qubits. We observe from Eq.~\eqref{eq:pz} that if the degree of dephasing can be reduced, MTQC can have a larger photon-loss threshold $\eta_{\rm th}$. To reduce the effect of dephasing, one intuitive approach would be to encode the qubits of $\ket{\mathcal{C}_\mathcal{L}}$ with a multiqubit repetition code.  This can be achieved by encoding the $m$-photon qubits of resource ket $\ket{\mathcal{C}_{3^\prime}}_{n,m,n}$ with an $N$-qubit repetition code; one would replace $\ket{0_m}$ with $\left(\ket{0_m}+\ket{1_m}\right)^{\otimes N} $, and  
$\ket{1_m}$ with $\left(\ket{0_m}-\ket{1_m}\right)^{\otimes N}$ (up to normalization), where $N$ is the repetition number. This gives the following encoded resource ket
\bea
\ket{\mathcal{C}_{3^\prime}}_{\rm enc}&=&\ket{0_n}\left(\ket{0_m}+\ket{1_m}\right)^{\otimes N}\ket{0_n}\nonumber\\
&&~~+\ket{1_n}\left(\ket{0_m}-\ket{1_m}\right)^{\otimes N}\ket{1_n}.
\label{eq:c3enc}
\eea
Note that the extreme qubits, each holding $n$ photons, are not encoded as they shall be consumed by $n$-BSMs anyway.
 
As an example for demonstrating that we can increase $\eta_{\rm th}$, let us consider $N=3$. The generation of the encoded $\ket{\mathcal{C}_{3^\prime}}$ kets using $\ket{\rm GHZ_3}$'s and ${\rm B_S}$ is explained in Appendix~\ref{app:newc3}. The  QEC procedure using this three-qubit repetition code employs the {\it majority voting} strategy, which fails when two of more qubits undergo dephasing~\cite{NC10}. Thus, the effective dephasing rate due to photon loss on the encoded lattice qubits is 
\be
p_{Z,\mathrm{enc}}=3p_Z^2(1-p_Z)+p_Z^3\,,
\label{eq:eff}
\ee
where $p_Z$ is the unencoded dephasing error from Eq.~\eqref{eq:pz}. It is clear that $p_{Z,\mathrm{enc}}<p_Z$ when $p_Z<0.5$. Hence, such a repetition encoding can suppress the dephasing rate, which would result in an improvement on $\eta_{\rm th}$.  The new tolerable photon-loss rate, $\eta_{\rm th}^{\rm enc}$ is deduced by inverting the expression for $p_{Z,\mathrm{enc}}$. In principle, we can make $\eta_{\rm th}^{\rm enc}$ arbitrarily close to 1 by increasing the value of $N$. This encoding strategy would evidently require more GHZ ingredient states to generate \emph{encoded} (star-)cluster states. However, as we shall see in Sec.~\ref{sec:resource}, numerical simulations show that using encoded states also reduces the effective dephasing rate (implying a larger tolerable photon-loss rate) to the extent that outweighs the additional GHZ states needed for encoding, such that smaller values of $d$ would suffice to reach $p_{\rm L}^{\rm targ}$.

We suppose that now, the lattice qubit is encoded with a \emph{finite} $N$-qubit repetition code \cite{NC10}. Using majority voting, the effective dephasing rate on the encoded lattice qubits is
\begin{equation}
	p_{Z,\mathrm{enc}}=\sum^N_{q=\lceil(N+1)/2\rceil}\binom{N}{q}p_Z^q(1-p_Z)^{N-q}\,,
	\label{eq:form1}
\end{equation} 
where $p_Z$ is the dephasing rate on un-encoded lattice qubits. The task is to reveal the influence of $N$ on the value of the photon-loss threshold rate 
$\eta_\mathrm{th}$.

We make use of the convenient approximation
\begin{equation}
	\binom{N}{q}p_Z^q(1-p_Z)^{N-q}\cong\dfrac{\exp\!\left(-\frac{(q-N p_Z)^2}{2\,Np_Z(1-p_Z)}\right)}{\sqrt{2\pi Np_Z(1-p_Z)}}
\end{equation}
that is valid for sufficiently large $N$ owing to the central limit theorem. After a variable substitution with $x=q/N$, this allows us to convert Eq.~\eqref{eq:form1} into an integral,
\begin{align}
	p_{Z,\mathrm{enc}}\cong&\,\int^\infty_{1/2}\D x\,\dfrac{\exp\!\left(-\frac{(x-p_Z)^2}{2\,p_Z(1-p_Z)/N}\right)}{\sqrt{2\pi p_Z(1-p_Z)/N}}\nonumber\\
	=&\,\dfrac{1}{2}-\dfrac{1}{2}\,\ERF{\dfrac{1-2p_Z}{2\sqrt{2p_Z(1-p_Z)/N}}}\,,
\end{align}
which involves the error function $\ERF{\cdot}$. The infinite upper limit of the integral is justified by the extremely narrow width of the Gaussian integrand for large $N$.
Since for a large argument $z$,
\begin{equation}
	\ERF{z}\cong1-\dfrac{\E{-z^2}}{\sqrt{\pi}\,z}\,,
\end{equation}
we get $y\,\E{y^2/8}\cong\sqrt{2/\pi}/p_{Z,\mathrm{enc}}$, where $y=\dfrac{1-2p_Z}{\sqrt{p_Z(1-p_Z)/N}}$. Squaring this relation allows us to write
\begin{equation}
	y^2\cong4\,\LambertW{1/\left(2\pi p^2_{Z,\mathrm{enc}}\right)}\,,
\end{equation}
which expresses the solution as a Lambert function~$\LambertW{\cdot}$. This leads to the physical solution
\begin{equation}
	p_Z\cong\dfrac{1}{2}-\dfrac{1}{2\sqrt{1+4\,t}}\,,\quad t=\dfrac{N}{4\,\LambertW{1/\left(2\pi p^2_{Z,\mathrm{enc}}\right)}}\,.
\end{equation}
We may now immediately identify $(1-\eta_\mathrm{th}^{\rm enc})^m\cong(1+4\,t)^{-1/2}$, with $m$ being the number of photons of each qubit in the repetition code, which finally yields
\begin{equation}
	\eta_\mathrm{th}^{\rm enc}\cong1-\left[1+\dfrac{N}{\LambertW{1/\left(2\pi p^2_{Z,\mathrm{enc}}\right)}}\right]^{-1/(2m)}\,.
\end{equation}

As $N$ increases, we find that $\eta_\mathrm{th}^{\rm enc}\cong 1-O(1/N^{1/(2m)})\rightarrow1$ for any designated value of $p_{Z,\mathrm{enc}}<p_Z$ and $m$. On the other hand, the function $\LambertW{1/\left(2\pi p^2_{Z,\mathrm{enc}}\right)}$ itself is a slowly increasing function of $p_{Z,\mathrm{enc}}$ as $p_{Z,\mathrm{enc}}$ decreases, originating from the large-argument expansion
\begin{align}
	\LambertW{x}\cong&\,\ln x-\ln \ln x+\frac{\ln \ln x}{\ln x}+\frac{(\ln \ln x-2) \ln \ln x}{2 (\ln x)^2}\,,
\end{align}
so that within the typical range $0.001\leq p_{Z,\mathrm{enc}}\leq0.1$ of interest, the order of magnitude for $\LambertW{1/\left(2\pi p^2_{Z,\mathrm{enc}}\right)}$ does not change. For a repetition code of fixed $N$ and a given $p_{Z,\mathrm{enc}}$, increasing $m$ reduces $\eta_\mathrm{th}^{\rm enc}$. 

\begin{figure*}
\includegraphics[width=1.6\columnwidth]{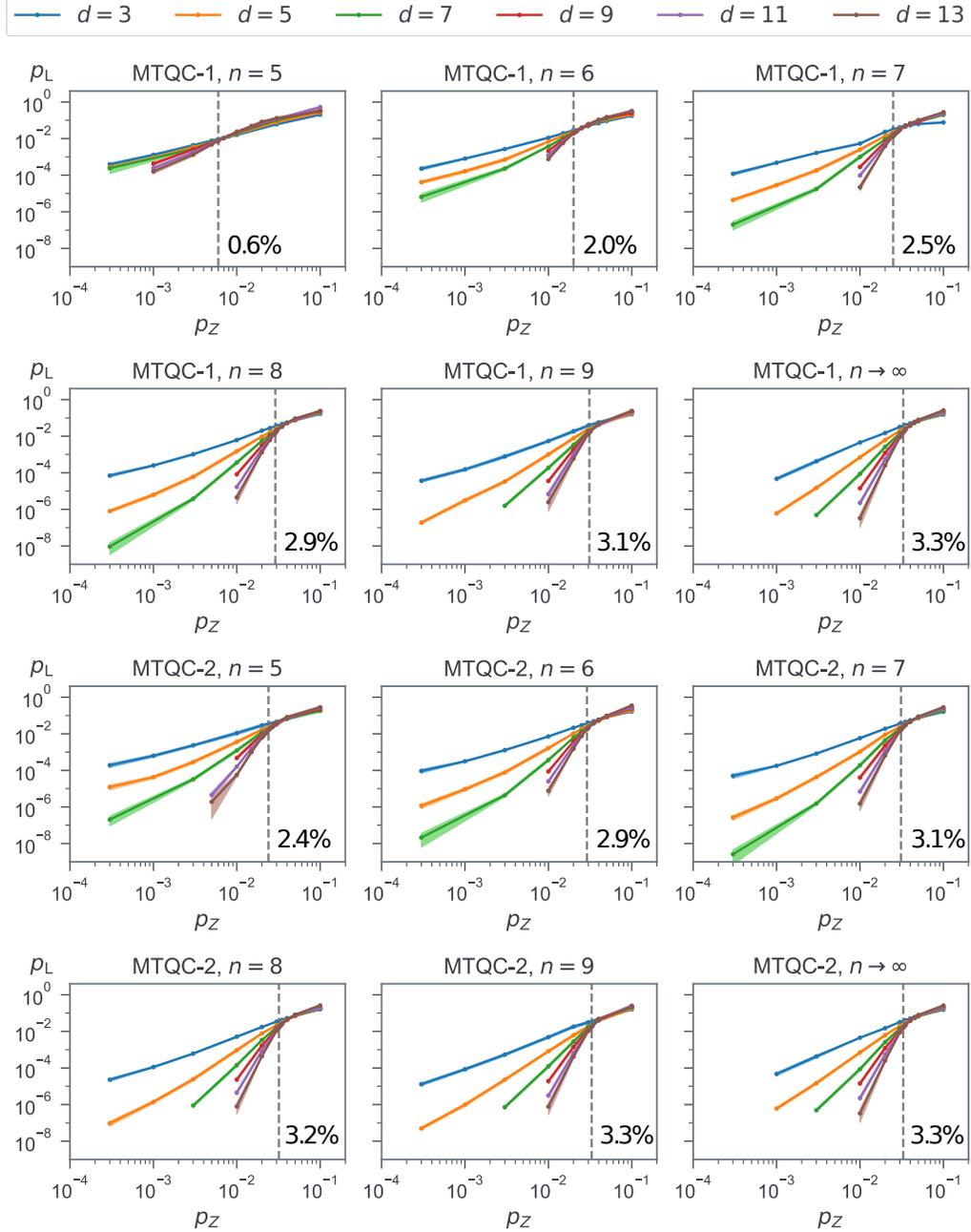}
\caption{Logical error rate $p_{\rm L}$ plotted against the dephasing rate $p_Z$ for MTQC-1 and MTQC-2 of $n\in[4,9]$, where $m=2$, accompanied by 99\% confidence intervals (shaded regions) that are typically much narrower than the average values. For each $n$ value, $p_{\rm L}$ corresponding to code distances (RHG lattice size) $d=3,5,7,9,11,13$ are plotted. The intersection point of the $p_{\rm L}$ curves for various $d$ values corresponds to the threshold dephasing rate $p_{Z,{\rm th}}$. We observe that as $n$ increases, the failure rate of $n$-BSMs, $p_{\rm f}$, decreases, leading to a larger $p_{Z,{\rm th}}$. It is important to note that when $n=9$, $p_{Z,{\rm th}}$ is close to that in the $p_{\rm f}=0$ case, so that considering $n>9$ results in no visible advantage. The threshold value $\eta_\mathrm{th}$ is shown for every figure panel.}
\label{fig:result}
\end{figure*}

\section{Results on photon-loss threshold}
\label{sec:result}
When carrying out QEC on $\ket{\mathcal{C}_\mathcal{L}}$, the logical error rate $p_{\rm L}$ is determined against $p_Z$ for various  code distances, $d$ via simulations. This procedure is repeated for various values of $n$, which determine the respective $p_{\rm f}$ of $n$-BSMs. The intersection point of the curves corresponding to various $d$'s is the threshold dephasing rate $p_{Z,\rm th}$ as marked in Fig.~\ref{fig:result}. The photon-loss threshold, $\eta_{\rm th}$ is determined using Eq.~\eqref{eq:pz} by replacing $p_Z$ with $p_{Z,\rm th}$. 

For example, from Fig.~\ref{fig:result} which corresponds to $n=8$, we have $p_{Z,{\rm th}}\approx2.9\times10^{-2}~(3.2\times10^{-2})$ for MTQC-1~\mbox{(MTQC-2)}. Considering that MTQC is operated below the threshold value (see Tab.~\ref{tab:result}) at $\eta=0.01$, according to Eq.~\eqref{eq:nBSM_fail}, we then have the associated value of $p_{\rm f}=4.58\times10^{-3}$. Replacing $p_Z$ by $p_{Z,{\rm th}}$ in Eq.~(\ref{eq:pz}) we find that the total tolerable photon loss rate, $\eta_{\rm th}$ is  
$2.9\times10^{-2}~(3.2\times10^{-2})$ for MTQC-1~(MTQC-2). However, the photon loss rate tolerable by individual components can be deduced as follows. The total time steps spent by lattice-qubits before being measured is $\kappa=3~(4)$ for MTQC-1~(MTQC-2). Therefore, we have $\eta_{\rm dly}=4.1\times10^{-3}(5.4\times10^{-3})$. Further, by inserting the value of $\eta_{\rm dly}$ in Eq.~\eqref{eq:comp_eta} ($m=2$), we have $\eta^{\rm soc}_{\rm th}=\eta^{\rm swc}_{\rm th}=\eta^{\rm det}_{\rm th}=8.6\times10^{-3}~(8.8\times10^{-3})$ for MTQC-1~(MTQC-2). This implies that equal amount of photon-loss can be tolerated in GHZ source, optical switches and measurement.
In this case each switch in the network can tolerate photon-loss rate of $\eta^{\rm s}=2.8\times10^{-3}(1.7\times10^{-3})$.
While the $\eta_{\rm th}$ in MTQC-2 is higher both subvariants offer comparable component-wise photon-loss thresholds. However, MTQC-2 imposes more stringent restriction on allowed photon-loss rate in switches. We lastly note that the time delays between consecutive layers can be neglected if they are sufficiently shorter than $-\left( L_0 / c \right) \ln \eta_\mathrm{th} \approx 4 \times 10^{-4}\text{ s}$ for both MTQC-1 and MTQC-2, as discussed in Sec.~\ref{sec:noise}.

The $\eta^{\rm soc}$, $\eta^{\rm swc}$ and $\eta^{\rm det}$ are complementary in nature as the overall tolerable photon-loss rate of the MTQC is fixed. Therefore, if the GHZ source and detectors are operated at lower loss levels, the MTQC protocol can tolerate a much higher photon-loss rate in the optical switches.
Similarly, we have repeated calculation for $n=5$ through~9 and the values of $\eta_{\rm th}$ are tabulated in Tab.~\ref{tab:result}. We know from the Ref.~\cite{AAG+18} that when $p_{\rm f}=0.145$ in MTQC-2, the $\ket{\mathcal{C}_\mathcal{L}}$ cannot tolerate any dephasing, and hence any photon loss. This threshold $p_{\rm f}$ is overcome when $n\geq3$. However, the situation is different for MTQC-1 and is explained in the following.

In MTQC-1, distorted $\ket{\mathcal{C}_\ast}$'s are allowed and this gives rise to  $\ket{\mathcal{C}_\mathcal{L}}$ with {\it diagonal} edges (refer also to Fig.~2(c) of Ref.~\cite{OTJ19}). This is overcome by removing the qubits at the ends of a diagonal edge during QEC. Therefore, the probability that a qubit survives on $\ket{\mathcal{C}_\mathcal{L}}$ is $1-p_{\rm f}$. Note also that each qubit in $\ket{\mathcal{C}_{\mathcal{L}}}$ is susceptible to losses from diagonal edged connected to \emph{four} $\ket{\mathcal{C}_*}$'s: two in the layer containing the qubit, and another two coming from different layers. Additionally, failure of an $n$-BSM that connects two $\ket{\mathcal{C}_\ast}$'s would leave an edge between qubits missing. This situation is handled by removing one of the qubits associated with such a missing edge~\cite{AAG+18}. The survival probability of a lattice qubit in this case is $1-p_{\rm f}/2$, where \emph{four} $n$-BSMs are involved. It follows that the total probability of loosing a qubit on $\ket{\mathcal{C}_\mathcal{L}}$ is $1-(1-p_{\rm f})^4(1-p_{\rm f}/2)^4$ and this number should not exceed $0.249$~\cite{LZ98,BS10} if $\ket{\mathcal{C}_\mathcal{L}}$ should be useful for quantum computing. Therefore, the threshold value for $p_{\rm f}$ is 0.047, which is determined by solving the equation $1-(1-p_{\rm f})^4(1-p_{\rm f}/2)^4=0.249$. So, MTQC-1 is possible only when $n\geq5$. Note that lattices other than the RHG type have different values for threshold failure rates of entangling operation~\cite{PTEG19,GSBR15}. Accordingly, the probability of missing qubits in MTQC-2 is $1-(1-p_{\rm f})^4$. For a given value of $p_{\rm f}$, the resulting lattice in MTQC-1 has more missing qubits and is therefore of a relatively poorer quality.

The tolerable photon-loss thresholds for MTQC naturally increase with the usage of  repetition codes in the manner discussed in Sec.~\ref{sec:qec}. Let us consider an example of encoded lattice-qubit with  $n=8$, $m=2$ and $N=3$. In this encoded situation, The value for the overall photon-loss threshold with a 3-qubit repetition code, obtained by using Eq.~(\ref{eq:eff}), is $\eta_{\rm th}^{\rm enc}=10.7\%~(11.1\%)$.   $\kappa=$6~(7) for MTQC-1~(MTQC-2), following which we have $\eta_{\rm dly}=8.1\times10^{-3}(9.5\times10^{-3})$.  The  component wise tolerence would be $3.2\times10^{-2}~(3.4\times10^{-2})$. In other words, encoding the lattice qubits with a three-qubit repetition code increases the $\eta_{\rm th}$ by nearly four times.
Also, each switch  can now tolerate a higher $\eta^{\rm s}=5.4\times10^{-3}~(4.9\times10^{-3})$. The results for other values of $n$ are available in Tab.~\ref{tab:result}.
Additionally, a sufficiently large set of stabilizer measurements on encoded qubits can be used to reconstruct the underlying noisy quantum process acting on these qubits~\cite{OSB15, OSB15s}.

\begin{table*}
\begin{tabular}{|c|c|c|c|c|c|c|c || c|c|c|c|c|c|c|c|} \hline
    &\multicolumn{7}{|c||}{MTQC-1}&  \multicolumn{7}{|c|}{MTQC-2}\\ \hline
$n$  & $p_{Z,{\rm th}}$ & $\eta_{\rm th}$ &  $\eta_{\rm th}^{\rm enc}$&$d_{10^{-6}}$&  $\mathcal{N}_{10^{-6}}$& $d_{10^{-6}}^{\rm enc}$ & $\mathcal{N}_{10^{-6}}^{{\rm enc}}$ & $p_{Z,{\rm th}}$ & $\eta_{\rm th}$& $\eta_{\rm th}^{\rm enc}$&$d_{10^{-6}}$ & $\mathcal{N}^{\,'}_{10^{-6}}$& $d_{10^{-6}}^{\rm enc}$ &$\mathcal{N}_{10^{-6}}^{\,'{\rm enc}}$\\ 
       &                      &                         &                         &$d_{10^{-15}}$ &$\mathcal{N}_{10^{-15}}$& $d_{10^{-15}}^{\rm enc}$   &$\mathcal{N}_{10^{-15}}^{{\rm enc}}$&                   &                        &                           &$d_{10^{-15}}$ &$\mathcal{N}^{\,'}_{10^{-15}}$& $d_{10^{-15}}^{\rm enc}$ & $\mathcal{N}_{10^{-15}}^{\,'{\rm enc}}$ \\ \hline
  \,\,\,5\,\,\,  & $0.006$ & $0.6\%$& $4.6\%$& 53      & $1.13\times10^9$& 39    &$7.40\times 10^8$ &  $0.024$ & $2.4\%$& $9.7\%$& 21 & $7.66\times 10^7$& 7 &$3.42\times 10^{6}$\\ 
   &&&& 168  & $3.66\times 10^{10}$ &    162  & $5.5\times 10^{10}$ &    &&& 61 & $1.91\times 10^{9}$& 22 & $1.47\times10^{8}$   \\ \hline
  6   &$0.02$ & $2.0\%$ &$8.7\%$ &  41   & $8.91\times 10^8$ &  9   &$1.47\times 10^7$ &  $0.029$ & $2.9\%$ & $10.7\%$&  17  & $5.90\times 10^7$& 5  &$2.42\times 10^6$\\ 
   &&&&  128   & $2.74\times 10^{10}$ &   34    & $6.82\times 10^{8}$ &    && & 51  &$1.76\times 10^{9}$& 15  & $5.66\times10^{7}$   \\ \hline
  7  & $0.025$ & $2.5\%$&$9.9\%$&  19     & $1.34\times 10^8$&   6   &$4.73\times 10^6$ &  $0.031$ & $3.0\%$& $10.9\%$& 15   & $5.58\times10^7$ & 5  &$2.13\times 10^7$\\ 
 &&&&  58     & $3.56\times10^9$ &   19    & $1.45\times 10^{8}$ &    &&& 42  &$1.29\times 10^9$& 13 & $4.46\times 10^7$  \\ \hline
  8   &0.029 & $2.9\%$& $10.7\%$ &  15      &$8.19\times10^7$&  5     &$3.28\times10^6$ &  $0.032$&  $3.1\%$   & $11.1\%$ &  13   & $4.78\times 10^7$ &  4  &$2.00\times10^6$\\ 
     &&&&  47    &$2.33\times 10^9$ &     14  &$7.94\times10^7$ &    &&&   37  &$1.14\times10^9$& 12   &$4.51\times10^7$ \\ \hline
  9   &  $0.031$ &  $3.1\%$&  $11.1\%$ &  14  & $8.25\times10^7$&  4  &$2.86\times10^6$ &  $0.033$ &  $3.3\%$&  $11.5\%$ & 12  &  $5.93\times10^7$ & 4  & $2.07\times10^6$\\ 
     &&&&   47   &$3.16\times 10^9$ &  12    &$6.28\times10^7$ &    &&&   35  &$1.28\times10^9$& 11   &$5.03\times10^7$   \\ \hline
\end{tabular}
\caption{Table of values for the dephasing-noise threshold $p_{Z,{\rm th}}$, photon-loss threshold $\eta_{\rm th}$ and resource overhead $\mathcal{N}^{(\,')}_{p_{\rm L}^{\rm targ}}$ to achieve $p^\mathrm{targ}_{\rm L}=10^{-6}$ and $10^{-15}$ concerning various instances $n$ of our scalable MTQC-1 and MTQC-2 protocols. The improvement in photon-loss threshold by encoding all lattice qubits with the ($N=3$)-qubit repetition QEC code, $\eta_{\rm th}^{\rm enc}$ and their associated resource overhead, $\mathcal{N}_{p_{\rm L}^{\rm targ}}^{(\,')\rm enc}$ are also listed for comparisons. The table starts from $n=5$ since MTQC-1 is not possible when $n\leq4$ as detailed in Sec.~\ref{sec:result}.  The benefits of  encoding is apparent both in terms of $\eta_{\rm th}$ and resource overheads. The asymptotically achievable value of $p_{\rm th}$ when  $p_{\rm f}=0$ is 0.033 (see Fig.~\ref{fig:result}).  Therefore, increasing $n$ beyond 9 gives no visible advantage. However, $\eta_{\rm th}^{\rm enc}$ can be arbitrarily improved by increasing the repetition number $N$ in the encoding of lattice-qubits. As is expected, increasing the value of $n$ generally improves $\eta_{\rm th}$ and $\eta_{\rm th}^{\rm enc}$. Interestingly, $\mathcal{N}_{p_{\rm L}^{\rm targ}}$ reduces with $n$ until $n=8$, beyond which the excessive amount of $\ket{{\rm GHZ}_3}$'s required for resource-states generation quickly nullifies any subsequently insignificant improvement in $\eta_{\rm th}$. In view of this, we conclude that MTQC-2 with $n=8$ is the optimal case of un-encoded MTQC. Note that the resource overheads are calculated when MTQC operates at $\eta=0.01$ which corresponds to $p_Z=0.01$ and $p_{Z,{\rm enc}}=0.0003$. A similar practice is adopted in~\cite{RHG07}, where ``operational overheads'' are computed ``at 1/3 of the fault-tolerance threshold.''}
\label{tab:result}
\end{table*}

\section{Results on Resource overhead} 
\label{sec:resource}
In this work, we consider GHZ states as the raw ingredients for constructing $\ket{\mathcal{C}}_\mathcal{L}$, on which fault-tolerant gate operations are performed via single-qubit measurements. Therefore, the resource overhead, $\mathcal{N}$, is the average number of $\ket{{\rm GHZ}}$'s consumed to build $\ket{\mathcal{C}}_\mathcal{L}$ of required size. Other components used in MTQC, such as delay lines, detectors, optical switches and beam splitters scale proportionately to $\mathcal{N}$. Alternatively, Ref.~\cite{LHMB15} considers detectors to be resources. Logical gate operations are possible by passing  defects through $\ket{\mathcal{C}}_\mathcal{L}$~\cite{RHG06,RHG07}. As logical errors occur when a chain of $Z$ errors connects two defects or encircles a defect, error-free operations would demand these defects be separated by a distance $d$ and also have a perimeter of $d$. When noise is below threshold, by increasing the value of $d$, the logical error rate $p_{\rm L}$ can be reduced arbitrarily. If $\ket{\mathcal{C}}_\mathcal{L}$ has sides of length $l=5d/4$, it can accommodate a defect of perimeter~$d$  and other defects placed a distance $d$ apart from each other (refer to Fig.~8 of Ref.~\cite{OTJ20}).

The time for simulation of QEC on $\ket{\mathcal{C}}_\mathcal{L}$  increases drastically with $d$, rendering the estimation of (a very small) $p_{\rm L}$ for arbitrarily large $d$ unfeasible. So, the value of $d$ at which an extremely small target $p_{\rm L}$~($p^{\mathrm{targ}}_{\rm L}$) is achieved can be estimated by extrapolating known values of $p_{\rm L}$. We can determine the target $d$ required to achieve $p^{\mathrm{targ}}_{\rm L}=10^{-6}$ and $10^{-15}$ using the following expression~\cite{WF14}
\be
p^{\mathrm{targ}}_{\rm L}=b\,\left(\dfrac{a}{b}\right)^{\displaystyle-(d-d_b)/2},
\label{eq:pl}
\ee
where $a$ and $b$ are the values of $p_{\rm L}$ corresponding to the second largest $d_a$ and the largest code distance $d_b$, respectively considered in our simulations. For example, as seen from the Fig.~\ref{fig:result} when $p_Z=0.01$ we have $d_a=11$ and $d_b=13$. 

As a practice, for $n>5$, we operate MTQC at $\eta=0.01$, which is below the photon-loss threshold value. This corresponds to $p_Z\approx0.01$ at $m=2$ for the unencoded case and $p_{Z,\mathrm{enc}}\approx 3\times10^{-4}$ for the encoded one. However, unencoded MTQC-1 with $n=5$ yields a threshold rate of $\eta_{\rm th}\approx6\times10^{-3}$ and is thus operated at $\eta=3\times10^{-3}$. Therefore, the $\mathcal{N}_{p_{\rm L}^{\rm targ}}$ is also determined at the operation point. In Eq.~(\ref{eq:pl}), $a$ and $b$ also correspond to the operation point. The reason for choosing the operation point away from the threshold is as follows: It is known empirically that $p_{\rm L}\propto (p_Z/p_{Z,{\rm th}})^{(d+1)/2}$ when the {\it minimum weight perfect matching} decoder is used~\cite{FMM+12}. If we operate closer to the threshold, a larger $d$ is essential to reach some pre-chosen $p_{\rm L}^{\rm targ}$. Thus, the operation point is chosen away from the threshold. Also, sufficiently away from the threshold point the ratio $a/b$ is reasonably constant and the estimation with Eq.~\eqref{eq:pl} is reliable~\cite{WF14}. Once $d$ for achieving $p^{\mathrm{targ}}_{\rm L}$ is determined, $\mathcal{N}$ can be estimated by counting the average number of GHZ states required to build  $\ket{\mathcal{C}}_\mathcal{L}$ of side $l=5d/4$. Only the central qubit of a  $\ket{\mathcal{C}_\ast}$ stays in the lattice and rest of them are consumed by $n$-BSMs. A $\ket{\mathcal{C}}_\mathcal{L}$ of sides $l$ would have $6l^3$ qubits. Therefore, we need $6l^3$  $\ket{\mathcal{C}_\ast}$'s per fault-tolerant gate operation.

As explained in Sec.~\ref{sec:star}, to create a $\ket{\mathcal{C}_\ast}$ we need two $\ket{\mathcal{C}_3}_{n,n,n}$'s and one $\ket{\mathcal{C}_{3^\prime}}_{n,m,n}$. Based on the noise model in Sec.~\ref{sec:noise}, in order to minimize dephasing effects on the central qubit, we set $m=2$. According to Fig.~\ref{fig:c3}, the creation of a $\ket{\mathcal{C}_3}_{n,n,n}$ necessitates the entanglement of two 
$\ket{{\rm GHZ}_{n+1}}$ and a $\ket{{\rm GHZ}_{n+2}}$ using two ${\rm B_S}$'s. 
As the failure rate of one ${\rm B_S}$ is $(1+2\eta)/2$ given a photon-loss rate $\eta$ [obtained by setting $n=1$ in Eq.~\eqref{eq:nBSM_fail}], one needs $8(1-2\eta)^{-2}$ copies of  $\ket{{\rm GHZ}_{n+1}}$ and $4(1-2\eta)^{-2}$ copies of $\ket{{\rm GHZ}_{n+2}}$, on average. Similarly, to create a $\ket{\mathcal{C}_{3^\prime}}_{n,m,n}$, one needs on average of $8(1-2\eta)^{-2}$ $\ket{{\rm GHZ}_{n+1}}$ and $4(1-2\eta)^{-2}$ $\ket{{\rm GHZ}_{m+2}}$.
Therefore, 
\be
\mathcal{N}_\ast=\dfrac{4(6N_{n+1}+2N_{n+2}+N_{m+2})}{(1-2\eta)^{2}}
\label{nstar1}
\ee
$\ket{{\rm GHZ}_3}$'s consumed to create a $\ket{\mathcal{C}_\ast}$ in MTQC-1, on average, where $N_r$ is the average number of $\ket{{\rm GHZ}_3}$'s consumed to create a $\ket{{\rm GHZ}_r}$, example values of which can be read from Tab.~\ref{tab:ghz} in Appendix~\ref{app:GHZ_r}. On the other hand, one needs  
\be
\mathcal{N}_\ast^\prime=\dfrac{4(6N_{n+1}+2N_{n+2}+N_{m+2})}{(1-2\eta)^{2}(1-p_{\rm f})^{2}}\label{nstar2}
\ee
$\ket{{\rm GHZ}_3}$'s, on average, in MTQC-2. For example, consider the case when $n=8$ and $m=2$, and  MTQC is operated at $\eta=0.01$.
Looking at Tab.~\ref{tab:ghz} and inserting the values of $N_{10}$, $N_9$ and $N_4$ in to Eq.~(\ref{nstar1}) one can estimate that $4(6\times 55.16+2\times 68.00+4.08)(1-2\times0.01)^{-2}\approx1962$ $\ket{{\rm GHZ}_3}$'s, on average, are consumed in MTQC-1 to create a $\ket{\mathcal{C}_\ast}$. Similarly, using Eq.~(\ref{nstar2}) we estimate that approximately $1980$ $\ket{{\rm GHZ}_3}$'s are needed in MTQC-2.

Once the average number of $\ket{{\rm GHZ}_3}$'s required to create a $\ket{\mathcal{C}_\ast}$ is known, it is straight forward to calculate the resource overhead $\mathcal{N}_{p^\mathrm{targ}_{\rm L}}$ to achieve some $p^\mathrm{targ}_{\rm L}$. In the case of MTQC-1, we have
\be
\mathcal{N}_{p_{\rm L}^{\rm targ}}=\frac{375(6N_{n+1}+2N_{n+2}+N_{m+2})}{8(1-2\eta)^2}d^3, 
\label{n1}
\ee
and for MTQC-2, it is
\be
\mathcal{N}_{p_{\rm L}^{\rm targ}}^\prime=\frac{375(6N_{n+1}+2N_{n+2}+N_{m+2})}{8(1-2\eta)^2(1-p_{\rm f})^2}d^3.
\label{n2}
\ee
Let us consider the same example when $n=8$, $m=2$ and similar value for $\eta$ to estimate resource overheads.  Further, from the simulation results we estimate that one needs $d\approx15~(13)$ to attain $p_{\rm L}\sim10^{-6}$ in MTQC-1~(MTQC-2).
Using the values of $d$ in Eqs.~(\ref{n1}) and (\ref{n2}) we estimate that   $\mathcal{N}_{10^{-6}}\approx8.19\times10^7$ and  $\mathcal{N}_{10^{-6}}^\prime\approx4.78\times10^7$.
On the other hand, for  $p_{\rm L}\sim10^{-15}$  one needs $d\approx47~(37)$ and thus  $\mathcal{N}_{10^{-15}}\approx2.33\times10^9$ and $\mathcal{N}_{10^{-15}}^\prime \approx1.14\times10^9$. Resource overheads and $d$ for other values of $n$ are presented in Tab.~\ref{tab:result}.

\begin{figure*}[t]
	\centering
	\includegraphics[width=2.1\columnwidth]{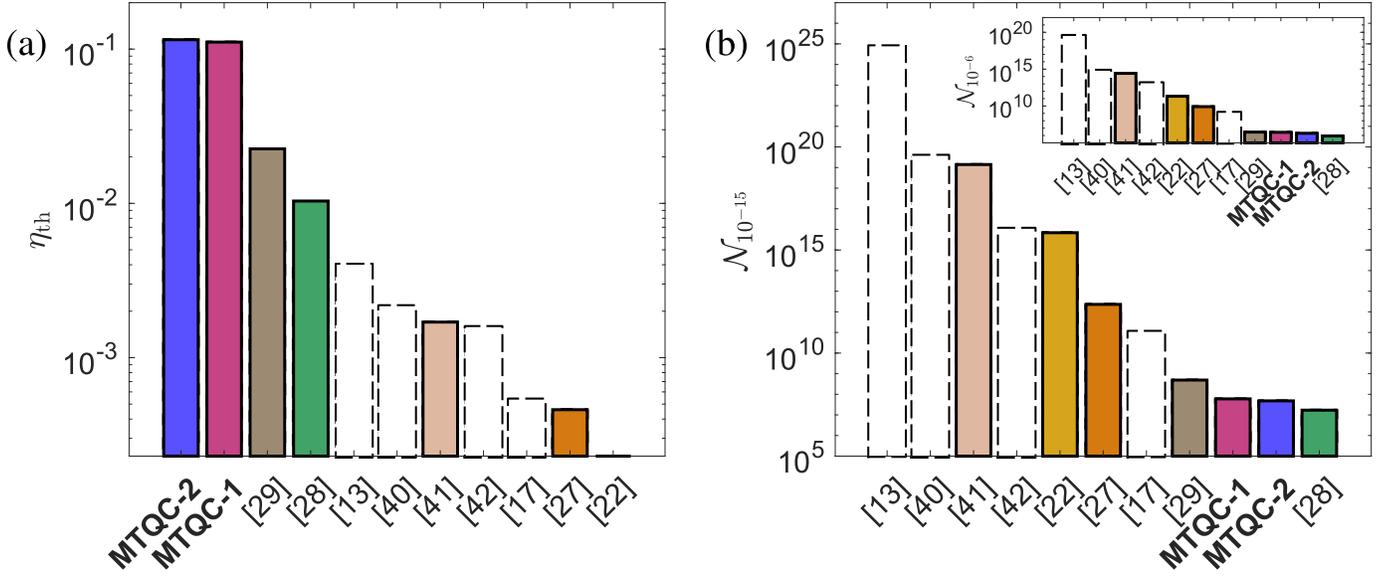}
	\caption{Contrast of the MTQC-1 and MTQC-2 paradigms, concatenated with repetition codes, against several other reported linear optical quantum computing schemes. (a) In terms of photon-loss threshold $\eta_{\rm th}$, we note that both MTQC-1 and MTQC-2 permit an overall tolerance value that is conspicuously larger than those offered in Refs.~\cite{DHN06,HHGR10,LPRJ15,Cho07} by at least an order of magnitude, and at least two orders of magnitude than those given in Refs.~\cite{HFJR10, LJ13, LRH08}. 
	The $\eta_\mathrm{th}$'s shown here for the schemes in Refs.~\cite{OTJ19,OTJ20} are at least three times as large as the originally-published values. This is because, here, we quote the \emph{overall} threshold values for a fair comparison with other schemes; original articles presented component-wise values. It should also be noted that the $\eta_{\rm th}$'s of schemes from Refs.~\cite{DHN06,HHGR10,Cho07, HFJR10}, represented by empty bars with dashed borders, are valid only for zero depolarizing error, which is physically unachievable; their loss thresholds under the equally realistic condition including depolarizing (or dephasing) errors must be much lower than the presented values.
		(b) Both MTQC schemes also excel in terms of resource overheads required to reach a target logical error rate $\mathcal{N}_{p_{\rm L}^{\rm targ}}$. The inset chart corresponds to  $\mathcal{N}_{10^{-6}}$, while the main chart one corresponds to $\mathcal{N}_{10^{-15}}$. It is apparent that MTQC is efficient in $\mathcal{N}_{p_{\rm L}^{\rm targ}}$ than other schemes (other than that in Ref.~\cite{OTJ19}) by several orders of magnitude. Importantly, MTQC ranks almost comparably with~\cite{OTJ19} as the most resource-efficient schemes known.}
	\label{fig:eta-res}
\end{figure*}  

In encoding the qubits of $\ket{\mathcal{C}}_\mathcal{L}$ with the three-qubit repetition QEC code  $\ket{\mathcal{C}_{3^\prime}}_{n,m,n}$ is replaced by  $\ket{\mathcal{C}_{3^\prime}}_{\rm enc}$ whic is created by replacing  $\ket{{\rm GHZ}_{m+2}}$ in Fig.~\ref{fig:c3} by 
\bea
\ket{\rm{enc}}&=&\ket{\textsc{h}}\left(\ket{\textsc{h}}^{\otimes m}+\ket{\textsc{v}}^{\otimes m}\right)^{\otimes3}\ket{\textsc{h}}\nonumber\\
&&~~+\ket{\textsc{v}}\left(\ket{\textsc{h}}^{\otimes m}-\ket{\textsc{v}}^{\otimes m}\right)^{\otimes3}\ket{\textsc{v}}.
\label{eq:enc}
\eea
The process to create $\ket{\rm{enc}}$ is detailed in Appendix~\ref{app:newc3}. To estimate the resource overhead due to encoding, we first estimate the average number of $\ket{{\rm GHZ}_{3}}$'s consumed to generate $\ket{{\rm enc}}$. For this a $\ket{{\rm GHZ}_{3}}$ and a $\ket{{\rm GHZ}_{5}}$ are needed in the first step (refer to Appendix~\ref{app:newc3}) and are entangled using ${\rm B_S}$.  When $\eta=0.01$, $N_{\rm enc}\approx104.96$ $\ket{{\rm GHZ}_{3}}$'s are incurred, on average, to form a $\ket{{\rm enc}}$  (deduced in Appendix~\ref{app:newc3}). 
Hereafter, the procedure to estimate resource overhead remains the same as non-encoded case.  Therefore, the resource overheads in the encoded case are
\be
\left.\mathcal{N}_{p^\mathrm{targ}_{\rm L}}^{{\rm enc}}\right|_{m=2}=\frac{375(6N_{n+1}+2N_{n+2}+N_{\rm enc})}{8(1-2\eta)^2}d^3,
\label{n1enc}
\ee
for MTQC-1 and
\be
\left.\mathcal{N}_{p^\mathrm{targ}_{\rm L}}^{\prime~{\rm enc}}\right|_{m=2}=\frac{375(6N_{n+1}+2N_{n+2}+N_{\rm enc})}{8(1-2\eta)^2(1-p_{\rm f})^2}d^3
\label{n2enc}
\ee
for MTQC-2.

Let us  re-estimate the resource overhead for the same example case with $n=8$, $m=2$ and $\eta=0.01$  with lattice qubits being encoded. As the encoded MTQC operating under similar photon-loss condition gives rise to smaller dephasing; that is, $p_{\rm eff}<p_Z$, smaller $d$ values suffice to attain $p_{\rm L}^{\rm targ}$. The same is refelected in the simulation results.
Now, from the simulation we estimate that $d\approx5~(4)$ is essential to attain $p_{\rm L}\sim10^{-6}$ in MTQC-1 (MTQC-2).
Inserting the values of $N_{10}$, $N_9$, $N_{\rm enc}$ and $\eta$ in to Eq.~(\ref{n1enc}) one can estimate that 
$\mathcal{N}_{10^{-6}}^{\rm enc}\approx3.28\times10^6$ and $\mathcal{N}_{10^{-6}}^{\prime~{\rm enc}}\approx2.0\times10^6$.
Similarly, to attain $p_{\rm L}\sim10^{-15}$ one needs  $d\approx14~(12)$. Therefore, $\mathcal{N}_{10^{-15}}^{\rm enc}\approx7.94\times10^7$ and $\mathcal{N}_{10^{-15}}^{\prime~{\rm enc}}\approx4.41\times10^7$. Resource overheads and $d$ in encoded case for other values of $n$ are presented in Tab.~\ref{tab:result}

\section{Comparison}
\label{sec:compare}

Now, we shall compare the performance of our MTQC to other schemes for fault-tolerant linear optical quantum computing. In Fig.~\ref{fig:eta-res}, we present  (a) photon-loss thresholds and (b) resource overheads of known linear optical quantum computing schemes \cite{OTJ20,DHN06,OTJ19,HHGR10,LPRJ15,Cho07,HFJR10, LJ13, LRH08} with MTQC-1 and MTQC-2. Clearly, MTQC shows exceptionally high loss tolerance compared to all known schemes, and is also highly competitive in terms of resource efficiency. In the following we shall briefly describe each scheme to which MTQC is compared.

Reference~\cite{DHN06} is one of the first works to determine the region of $\eta_{\rm th}$ along with a dephasing error rate and an estimation of resource overheads for fault-tolerant linear optical quantum computing.  The scheme uses optical cluster states built using polarization Bell pairs. 
This scheme couples 7-qubit Steane QEC codes~\cite{NC10} with telecorrection (where teleportation is used for error-syndrome extraction) for fault-tolerance. Unlike schemes that use topological codes, the  concatenation of Calderbank--Shor--Steane~(CSS) codes with itself is employed to attain smaller values of $p_{\rm L}$. For example, four~(six) levels of concatenation were employed to achieve $p_{\rm L}\sim10^{-6}~(10^{-15})$. When $\eta=4\times10^{-3}$ and depolarization rate is $4\times10^{-3}$, one has $\mathcal{N}_{10^{-6}}\approx2.6\times10^{19}$ and $\mathcal{N}_{10^{-15}}\approx7.1\times10^{24}$. The resource overhead demanded by the scheme is too high for practical considerations. A subsequent scheme in Ref.~\cite{HHGR10} that encodes multiple polarization photons into a logical qubit in a {\it parity state} provides a smaller $\eta_{\rm th}\approx2\times10^{-3}$, but an improved resource efficiency compared to Ref.~\cite{DHN06}. This scheme also uses 7-qubit Steane QEC codes with telecorrection and multiple levels of concatenations. When $\eta=4\times10^{-3}$ and depolarization rate is $4\times10^{-3}$, the average number of Bell pairs consumed is $\mathcal{N}_{10^{-6}}\approx6.8\times10^{14}$ and $\mathcal{N}_{10^{-15}}\approx3.5\times10^{19}$~\cite{OTJ20} with four and six levels of concatenation, respectively. 

Later, Ref.~\cite{Cho07} used {\it error-detecting quantum state transfer} where the underlying codes were capable of detecting errors in a way similar to the scheme in Ref.~\cite{Kni05}. Here, QEC is done by concatenating different error-detecting codes.
This scheme offers a smaller $\eta_{\rm th}\approx1.57\times10^{-3}$, but the value of $N$ could be reduced by many orders of magnitude compared to Ref.~\cite{DHN06}. 
When $\eta=1\times10^{-4}$ and depolarizing rate is $1\times10^{-5}$, with five~(seven) levels of concatenation, the average number of Bell pairs used is $\mathcal{N}_{10^{-6}}\approx O(10^{13})~[\mathcal{N}_{10^{-15}}\approx O(10^{16})]$. 
There is yet another multi-polarization-photon qubit quantum computing scheme~\cite{LPRJ15} that again
utilizes telecorrection based on 7-qubit Steane QEC codes and thus needs need the same levels of concatenation. Calculations in Ref.~\cite{OTJ19} show that   $\mathcal{N}_{10^{-6}}\approx O(10^{13})~[\mathcal{N}_{10^{-15}}\approx O(10^{16})]$ Bell pairs are consumed when $\eta\approx O(10^{-4})$ and $p_Z\approx O(10^{-4})$.

Using streams of entangled polarization photons, a topological photonic quantum computing scheme, which involves creating $\ket{\mathcal{C}_\mathcal{L}}$, was proposed in Ref.~\cite{HFJR10}. This has $\eta_{\rm th}\approx5.3\times10^{-4}$ when the depolarizing error rate is zero. However, this scheme gives $p_{\rm th}=1.14\times10^{-3}$ for the hypothetical case of $\eta=0$, which is higher by an order of magnitude compared to other non-topological fault-tolerant architectures. Calculations in Ref.~\cite{OTJ19} show that  $\mathcal{N}_{10^{-6}}>2\times10^{9}~(\mathcal{N}_{10^{-15}}>4.2\times10^{10})$ for non-zero $\eta$. 

The coherent-states $\{\ket{\alpha},\ket{-\alpha}\}$ can be used as the logical basis for CV qubits~\cite{JK02,Ralph03,LRH08}. Reference~\cite{LRH08} uses these qubits to develop a fault-tolerant  quantum computing scheme. This also employs 7-qubit Steane QEC codes with telecorrection and multiple levels of concatenations for tolerance against photon-loss and dephasing errors. Here, superpositions of coherent states, $\ket{\alpha}\pm \ket{-\alpha}$ (up to normalization) \cite{YS86,Ourj07}, are considered as resources. For this scheme $\eta_{\rm th}\approx2.3\times10^{-4}$ and $\mathcal{N}_{10^{-6}}\approx2.1\times10^{11}~(\mathcal{N}_{10^{-15}}\approx6.9\times10^{15})$ when $\eta=8\times10^{-5}$ and $p_Z=2\times10^{-4}$. The resource overhead is reduced by many orders of magnitude compared to Ref.~\cite{DHN06}, but this comes at the cost of a very low $\eta_{\rm th}$. Reference~\cite{LJ13} improved this situation by replacing coherent superposition states with hybrid states. The new scheme offers a better value of $\eta_{\rm th}\approx4.6\times10^{-4}$. Here $\mathcal{N}_{10^{-6}}\approx8.2\times10^{9}~(\mathcal{N}_{10^{-15}}\approx2.3\times10^{12})$ hybrid qubits are required when $\eta=O(10^{-4})$ and $p_Z=O(10^{-4})$.

By far, Ref.~\cite{OTJ19} shows the best $\eta_{\rm th}$-to-resource-overhead ratio that is achievable by creating $\ket{\mathcal{C}_\mathcal{L}}$'s with hybrid qubits. In the scheme of Ref.~\cite{OTJ19}, $\eta_{\rm th}\approx3.3\times10^{-3}$ and $\mathcal{N}_{10^{-6}}\approx8.5\times10^{5}~(\mathcal{N}_{10^{-15}}\approx1.7\times10^{7})$ hybrid qubits are consumed when $\eta=1.5\times10^{-3}$ and $p_Z=3\times10^{-3}$. 
Subsequently, Ref.~\cite{OTJ20} demonstrated that $\eta_{\rm th}$ can be further improved by spending more hybrid qubits. This scheme could achieve an improved $\eta_{\rm th}\approx5.7\times10^{-3}$ with  $\mathcal{N}_{10^{-6}}\approx2.9\times10^{7}~(\mathcal{N}_{10^{-15}}\approx4.9\times10^{8})$ when $\eta=2.6\times10^{-3}$ and $p_Z=2.3\times10^{-3}$. 

After the previous overview of various existing linear optical schemes for fault-tolerant quantum computing and mentioning the associated numerical values of $\eta_{\rm th}$ and $\mathcal{N}_{p_{\rm L}^{\rm targ}}$, we are now set for a comparison with MTQC. We shall compare with schemes that involve the creation of RHG lattices---the schemes in Refs.~\cite{HFJR10,OTJ19,OTJ20}. In comparison, MTQC outperforms the scheme in Ref.~\cite{HFJR10} both in terms of photon-loss tolerance and resource efficiency. Although the  MTQC performs better than schemes in Refs.~\cite{OTJ19,OTJ20} in terms of $\eta_{\rm th}$ it falls short in terms of $\mathcal{N}_{p_{\rm L}^{\rm targ}}$ compared to the scheme in Ref.~\cite{OTJ19}.
On the other hand, the MTQC outperforms all the known non-topological schemes Refs.~\cite{HFJR10,DHN06,HHGR10,LPRJ15,Cho07,LRH08,LJ13} both in terms of $\eta_{\rm th}$ and resource efficiency.

We note that the error models in Refs.~\cite{DHN06,HHGR10,Cho07,HFJR10} employed photon loss and depolarizing errors independently. On the other hand, the other schemes 
\cite{OTJ20,OTJ19,LPRJ15,LJ13, LRH08} considered more realistic models where photon loss and dephasing are related. In addition, the photon-loss thresholds in Refs.~\cite{DHN06,HHGR10,Cho07,HFJR10} are obtained under the condition of zero depolarizing error, which is unrealistic. For these four schemes, when a non-zero depolarizing error is considered, the threshold values should then be lower than the ones presented as empty bars in Fig.~\ref{fig:eta-res}(a), since additional depolarizing errors deteriorate $\eta_\mathrm{th}$. 

A very recent scheme~\cite{FBQC} that also encodes lattice qubits with QEC codes claims to be able to tolerate a photon-loss rate of 10.4\% occurring in entangling operations. This requires a particular type of 24-photon entangled states of which information concerning their (unreported) generation resource overheads could be of interest. On the contrary, for MTQC in this work, we begin only with 3-photon GHZ states that can be deterministically generated using current technology \cite{Besse20}. 

The scheme in Ref.~\cite{LHMB15} considers detectors as resources, about $O(10^9)$ of them, to create a tree-cluster state, a collection of which ultimately forms a lattice~(similar to the fate of $\ket{\mathcal{C}_\ast}$'s). This can tolerate, component-wise, a photon-loss rate of approximately $1\times10^{-3}$ (when the beam splitters are assumed to be lossless and the success rate of a BSM is 0.5). In our work, when $n=9$, we need $2935$ $\ket{{\rm GHZ_3}}$'s~(encoded case) are consumed and this number also reflects the order of magnitude of detectors needed. Additionally, our protocol has component-wise $\eta_{\rm th}^{\rm enc}\approx3.7\times10^{-2}$, which is a significant improvement. The extravagant resource overhead originates from the usage of single photons as basic ingredients for forming $\ket{{\rm GHZ_3}}$'s (which are in turn used to generate tree-cluster states) with a low success rate of $1/32$~\cite{VBR08}. However, recent work has demonstrated that the success rate can be greater than $1/32$~\cite{Psi21}.

\section{Alternative platforms for scalability enhancement}
\label{sec:TF}

While fault-tolerant MTQC can significantly improve resource overheads and error thresholds relative to other schemes, it requires resource-state generation and moderately-large collective BSMs that could pose a challenge with current polarization-based optical platforms. In particular, repeat-until-success strategies based on polarization-encoded qubits, as discussed in this work and many of the cited references rely on repeated generation of resource states that involve a huge number of entangled multiphoton qubits. 

The use of time-bin qubits~(photons encoded into time-delayed pulses of well-separated arrival times that do not overlap one another~\cite{Morrison:2022frequency-bin}), which are alternative quantum-information encoding schemes, have been considered in the generation of multiphoton entangled states~\cite{Menicucci:2010arbitrarily,Lee_2019}. In particular, it has been shown that such an encoding allows for a deterministic generation of GHZ states~\cite{Besse:2020realizing}. In this reference, specifically, $\ket{\mathcal{C}_{3'}}_{n,m,n}$ of $2n+m=3$ and 4 were generated with the respective fidelity of~0.90 and~0.82. Improvement in the fidelity of larger time-bin GHZ states in order to generate $\ket{\mathcal{C}_{3'}}_{n,m,n}$ of $m\geq2$ and $n\geq8$, which are optimal ranges for MTQC as shown in Tab.~\ref{tab:result}, is an important research direction.

In the grander scheme of things, it might be of interest to consider the generation of both $\ket{\mathcal{C}_{3}}_{n,n,n}$ and $\ket{\mathcal{C}_{3'}}_{n,m,n}$ resource states using the full potential of temporal-mode~(TM) optical qubits with both time and frequency content~\cite{Brecht:2015photon,Humphreys:2014continuous-variable,Rafael:2016one-way,Roztocki:2020designing,Zhu:2021hypercubic}, where each mode possesses an infinite-dimensional Hilbert space that is encoded onto one physical qubit. By exploiting such a large number of degrees of freedom, it is, in principle, possible to encode multiqubit quantum information onto a single physical photon.

Let us briefly highlight how such a logical TM encoding works. The central operation is the (unitary) quantum pulse gate~(QPG)~\cite{Eckstein:2011quantum,Brecht:2011quantum,Brecht:2015photon,Gil-Lopez:2021universal}:
\begin{align}
    Q^{(\theta)}_{k}=&\,1-\ket{A_k}\bra{A_k}-\ket{C}\bra{C}+\cos\theta\,(\ket{A_k}\bra{A_k}+\ket{C}\bra{C})\nonumber\\
    &\,+\sin\theta\,(\ket{C}\bra{A_k}-\ket{A_k}\bra{C})\,,
\end{align}
which allows one to convert a mode-matched TM basis ket $\ket{A_k}$ ($\inner{A_k}{A_{k'}}=\delta_{k,k'}$) into the superposition $\ket{A_k}\cos\theta+\ket{C}\sin\theta$, where $\ket{C}$ is a TM in a different frequency band than $\ket{A_k}$ such that $\inner{A_k}{C}=0$. On the other hand, a mode-mismatched action, $Q^{(\theta)}_k\ket{A_{k'\neq k}}=\ket{A_{k'}}$, leaves the TM basis ket intact. 

To present an alternative approach to multiqubit cluster-state generation, just as an example, we shall revisit the so-called type-I fusion scheme~\cite{Browne:2005resource} that was first introduced to entangle two spatial photons using a PBS, followed by a photodetection after a $45^{\circ}$ polarization rotation. If we now suppose that $\ket{A_0}_a$ logically represents $\ket{0_n0_m0_n}$ for qubit~a and $\ket{A_1}_b$ logically represents $\ket{1_n1_m1_n}$ for qubit~b, then, it has already been shown in~\cite{Brecht:2015photon} that a QPG-adapted type-I fusion together with deterministic spatial-mode combination (implicitly carried out throughout this analysis) permits the generation of $\ket{\mathcal{C}_{3'}}_{\rm a}\equiv(\ket{A_0}_{\rm a}+\ket{A_1}_{\rm a})/\sqrt{2}$: 
\begin{align}
    &\,\ket{A_0}_\mathrm{a}\ket{A_1}_\mathrm{b}\xrightarrow{\displaystyle Q^{(\pi/4)}_{0,{\rm a}}Q^{(\pi/4)}_{1,{\rm b}}}(\ket{A_0}_\mathrm{a}+\ket{C}_\mathrm{a})(\ket{A_1}_\mathrm{b}+\ket{C}_\mathrm{b})\dfrac{1}{2}\nonumber\\
    &\,\xrightarrow[\displaystyle\text{single-photon heralding}]{\displaystyle\text{50:50 beam splitter on $C$ modes}}\ket{\mathcal{C}_{3'}}_{\rm a}\,.
    \label{eq:fuse1}
\end{align}
Here, single-photon heralding is performed consistently on \emph{one} of the two output detectors in order to fix all relative phase factors in the final output pure state. One may similarly continue the above type-I fusion procedure until four logical TM basis kets are superposed, resulting in the formation of $\ket{\mathcal{C}_{3}}_{\rm a}\equiv(\ket{A'_0}_{\rm a}+\ket{A'_1}_{\rm a}+\ket{A'_2}_{\rm a}-\ket{A'_3}_{\rm a})/2$:\\
\noindent
If $\ket{\psi_M}=\sum^{M-1}_{l=0}\ket{A'_l}/\sqrt{M}$, then
\begin{align}
    &\,\ket{\psi_2}\ket{A'_2}_\mathrm{a}\xrightarrow[\displaystyle Q^{(\tan^{-1}\sqrt{2})}_{2,{\rm b}}]{\displaystyle Q^{(-\pi/2)}_{1,{\rm a}}Q^{(\pi/4)}_{0,{\rm a}}}(\ket{\psi_2}_\mathrm{a}-\ket{C'}_\mathrm{a})\dfrac{1}{\sqrt{2}}\nonumber\\[-3ex]
    &\qquad\qquad\qquad\qquad\qquad\,\,\,\otimes(\ket{A'_2}_\mathrm{b}+\ket{C'}_\mathrm{b}\sqrt{2})\dfrac{1}{\sqrt{3}}\nonumber\\
    &\,\xrightarrow[\displaystyle\text{single-photon heralding}]{\displaystyle\text{50:50 beam splitter on $C'$ modes}}\ket{\psi_3}\,,
    \label{eq:fuse2}
\end{align}
and finally,
\begin{align}
    &\,\ket{\psi_3}_\mathrm{a}\ket{A'_3}_\mathrm{b}\xrightarrow[\displaystyle Q^{(\pi/3)}_{3,{\rm b}}]{\displaystyle Q^{(-\pi/12)}_{2,{\rm a}}Q^{(-\pi/2)}_{1,{\rm a}}Q^{(\pi/4)}_{0,{\rm a}}}(\ket{\psi_3}_\mathrm{a}-\ket{C'}_\mathrm{a})\dfrac{1}{\sqrt{2}}\nonumber\\[-3ex]
    &\qquad\qquad\qquad\qquad\qquad\qquad\qquad\,\,\,\otimes(\ket{A'_3}_\mathrm{b}+\ket{C'}_\mathrm{b}\sqrt{3})\dfrac{1}{2}\nonumber\\
    &\,\xrightarrow[\displaystyle\text{single-photon heralding}]{\displaystyle\text{50:50 beam splitter on $C'$ modes}}\ket{\mathcal{C}_3}_\mathrm{a}\,.
    \label{eq:fuse3}
\end{align}

\begin{figure}[t]
	\centering
	\includegraphics[width=0.6\columnwidth]{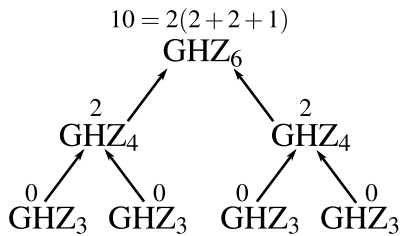}
	\caption{Counting the average number of PPOs to generate a GHZ$_6$ state from GHZ$_3$ states, labeled for every component state.}
	\label{fig:PPO}
\end{figure}

Ideally, near-perfect TM manipulation such as the above exemplifying scheme could reduce the average number of photon-pair operations~(PPOs) needed to generate resource states, since only two photons are handled at any round of the TM type-I fusion process \emph{via} a 50:50 beam splitter on the $C$ or $C'$ modes followed by detector-specific single-photon heralding. In practical scenarios, reducing the number of PPOs is equivalent to minimizing the number of detectors needed in a scheme as both scale commensurately with each other. As a basic comparison under ideal non-lossy conditions ($\eta=1$), and hence a 50\% chance of a BSM failure, we recall from Sec.~\ref{sec:resource} and Fig.~\ref{fig:c3} that two~GHZ$_{n+1}$ states and one~GHZ$_{m+2}$ state are entangled \emph{via} two BSMs to create a copy of $\ket{\mathcal{C}_{3'}}_{n,m,n}$ with polarization encoding. If we suppose that $m=2$ and $n=8$ are the target indices, then from Appendix~\ref{app:GHZ_r} and the fact that the average number of PPOs needed to create two states and entangle them, given that these states were previously generated with the respective average number of PPOs $l_1$ and $l_2$, is $2(l_1+l_2+1)$ in view of the 0.5 BSM failure rate, we require an average of 2~PPOs (\emph{via} BSMs) to create a GHZ$_4$ state and 34~PPOs to create a GHZ$_9$ state. These numbers are obtained from the assumption that no PPOs are required to generate the basic GHZ$_3$ states (see Fig.~\ref{fig:PPO} for a simple exposition). Therefore, an average of 218~PPOs are necessary to create a copy of $\ket{\mathcal{C}_{3'}}_{8,2,8}$. On the other hand, according to~\eqref{eq:fuse1}, TM-adapted type-I fusions only require 4~PPOs on average to create any $\ket{\mathcal{C}_{3'}}_{n,m,n}$ since the success probability of beam-split single-photon heralding on a specified detector is 0.25. Similarly, to create a copy of $\ket{\mathcal{C}_3}_{n,n,n}$, two~GHZ$_{n+1}$ states and one~GHZ$_{n+2}$ state are entangled. Repeating the above exercise, we find that a total of 378~PPOs are necessary to create $\ket{\mathcal{C}_3}_{8,8,8}$ using polarization encoding, whereas 64~PPOs with TMs [based on \eqref{eq:fuse1}--\eqref{eq:fuse3}] are sufficient to create any $\ket{\mathcal{C}_3}_{n,n,n}$.

The above comparison serves only to give a flavor of what TM encodings can do in terms of reducing the number of PPOs or detectors involved in resource-state generation. More accurate resource-overhead calculations are only available when detailed noise models and QPG mechanisms enter the analyses, requiring studies that are beyond the scope of this work. We remind the Reader that the above arguments are physically relevant \emph{provided that} orthogonal multiqubit information can be encoded into higher-order orthogonal TMs with high fidelity. However, realistic limitations on the bandwidth, photon-loss tolerance, TM shape switching speed and other imperfections that affect the QPG's output fidelity are the primary obstacles that prevent the generation of large superpositions at this stage. Additionally, appropriate error models affecting these TM states require careful and systematic analyses, together with the development of decoders suitable for this optical platform, are all crucial steps that shall be reserved for future studies. So, while the existing literature did pave the way for TM quantum computation, much more work is needed in order for practical applications to come to fruition.

\section{Discussion and Conclusion}
\label{sec:conclusion}
The work is motivated  by the recent advancements in experimental front of generation of deterministic multiphoton (polarization) entangled states like GHZ states \cite{Besse20}.
Here, we described an all-optical protocol that processes multiphoton GHZ states from deterministic sources using only passive linear-optical elements like beam splitter, delay lines, optical switches and only on-off detectors (no need for photon number resolution) to build RHG lattice for fault-tolerant quantum computing.  Major short comings of using polarization photons, the probabilistic entangling operation, is overcome by first creating multiphoton resource states and then performing $n$-Bell-state-measurements which are near-deterministic. However, the multiphoton resources states are created by entangling the three-photon GHZ states using probabilistic direct Bell-state-measurements.

Photon loss being major sources of errors, we demonstrated that our protocol  offers, without any quantum error correcting code concatenation, highest photon threshold of $3.1\%$ in MTQC-1  and $3.3\%$ in MTQC-2. Further, the photon-loss threshold is improved by encoding the lattice-qubits in three-qubit repetition code. This concatenation improves the threshold to $11.1\%$ in MTQC-1  and to $11.5\%$ in MTQC-2. We stress that this drastic improvement is made only with 3-photon GHZ states, passive linear-optical elements, and on-off detectors.
We also demonstrated that by employing codes of larger repetition number for concatenation, the photon-loss threshold can be further improved. Further, resource overheads in terms of the average number of three-photon GHZ states incurred per gate operation corresponding to various values of $n$ are tabulated in Tab.~\ref{tab:result}. In MTQC-2, when $n=9$ the resource overhead to reach the target logical error rate of $10^{-6}$~($10^{-15}$) is $2.07\times10^6~(5.03\times10^7)$. This is the most resource efficient case of MTQC. Interestingly, we observed that code concatenation not only improves photon-loss threshold but also favourably reduces the resource overheads of the MTQC. Comparing our results with known linear optical quantum computing schemes~\cite{HFJR10,DHN06,Cho07,HHGR10,LPRJ15,LRH08,LJ13,OTJ19,OTJ20}, MTQC offers clearly the highest tolerance against photon loss. MTQC is also highly resource-efficient compared with known linear optical schemes and is comparable only with Ref.~\cite{OTJ19}. In principle, our protocol can  be carried out even if we start with single photons as the basic ingredient. For this, the three-photon GHZ states can be generated using linear optics with a success rate of 1/32~\cite{VBR08}, or higher \cite{bartolucci2021creation}. In this case, the resource overhead (average number of single photons) for MTQC would increase approximately by two orders of magnitude.    

In the current work we have not used photon-number resolving detectors at any stage of the protocol that is essential to boost the success rate of direct BSM. The reason for this choice is that the on-off detectors are practically more efficient and can operate at room temperatures. If one chooses to employ photon-number resolving detectors and operate at cryogenic temperatures, the resource efficiency of MTQC can be further improved. 

Our protocol can also be extended to the creation of lattices with different geometry~\cite{SH21,Ben16}. However, it remains to be examined if they can be made tolerant against entangling operation failures.
MTQC also demonstrates the crucial need for the experimental development  of high fidelity deterministic multiphoton  entangled state generators in order for the advancement of the field of the scalable linear-optics-based quantum information processing.
Given its significant enhancement in the photon-loss threshold and the recent progress in generating multiphoton entanglement, our scheme will make scalable photonic quantum computing a step closer to reality.

\begin{acknowledgments}
This work was supported by National Research Foundation of Korea (NRF) grants funded by the Korea government (Grant Nos.~NRF-2019M3E4A1080074, NRF-2020R1A2C1008609 NRF-2020K2A9A1A06102946, NRF-2019R1A6A1A10073437 and NRF-2022M3E4A1076099) \emph{via} the Institute of Applied Physics at Seoul National University, and by the Institute of Information \& Communications Technology Planning \& Evaluation (IITP) grant funded by the Korea government~(MSIT) (IITP-2021-2020-0-01606, IITP-2021-0-01059). S.W.L. acknowledges support from the National Research Foundation of Korea (2020M3E4A1079939) and the KIST institutional program (2E31531). We thank Kamil~Bradler, Brendan~Pankovich, Alex~Neville, Jano~Gil-Lopez and Benjamin Brecht for insightful discussions. 
\end{acknowledgments}

\appendix

\section{Simulation of QEC}
\label{sec:sim}

\begin{figure}[t!]
	\centering
	\includegraphics[width=\columnwidth]{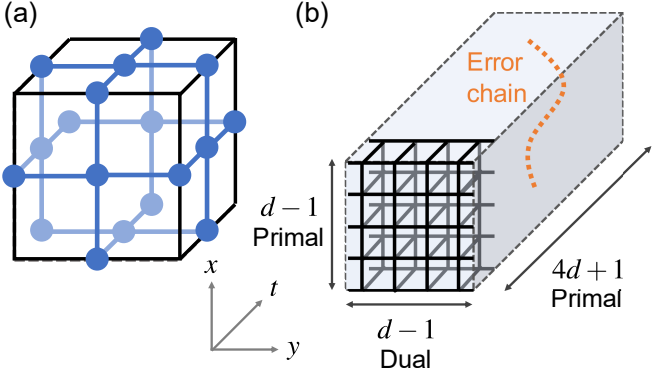}
	\caption{
		(a) Primal unit cell of an RHG lattice.
		Blue dots and lines indicate qubits and edges, respectively.
		(b) RHG lattice used for the simulation.
		The lattice is in a cuboid-shaped space with the size of $d-1$, $d-1$, and $4d+1$ along the $x$, $y$, and $t$ (simulating time) axes, respectively, in the unit of a cell.
		The first primal cells along the $t$ axis are shown as black solid lines.
		The boundaries are primal about the $x$ and $t$ axes, while dual about the $y$ axis.
		An error chain connecting the two opposite $x$ or $y$ boundaries (orange dotted line) incurs a logical error.
	}
	\label{fig:lattice_simulation}
\end{figure}

Here we present the method to obtain the logical error rate $p_\mathrm{L}$ numerically for a given subvariant of MTQC (MTQC-1 or MTQC-2), failure rate of $n$-BSM ($p_\mathrm{f}$), dephasing rate ($p_Z$), and code distance ($d$).
We consider a three-dimensional space with the $x$, $y$, and \textit{simulating time} ($t$) axes.
We simulate an RHG lattice in a cuboid with the size of ($d-1$, $d-1$, $T$) for $T:=4d+1$ about the three axes in the unit of a cell, as shown in Fig.~\ref{fig:lattice_simulation}.
The boundaries are primal about the $x$ and $t$ axes (that is, they are in contact with primal cells), while dual about the $y$ axis (that is, they cut primal cells in a half).

An isolated dephasing error is detected by two \textit{check operators} adjacent to the qubit.
Generally, an \textit{error chain} of dephasing errors is detected by two check operators located at its ends \cite{RHG06, RH07, RHG07, fowler2009topological}.
However, error chains connecting two opposite boundaries are not detectable, since there are no check operators at their ends.
If the number of such error chains regarding the $x$($y$)-boundaries is odd, a \textit{primal (dual) logical error} occurs.
We take account of only primal logical errors in this simulation.

The code distance is determined by the widths about the $x$ and $y$ axes, not the $t$ axis, thus $T$ can be an arbitrary number.
For fair comparison with different code distances or other computation schemes, we calculate the logical error rate \textit{per unit simulating time}.
$T$ should be large enough to get a reliable value, thus we set it to $4d+1$.

We use the Monte Carlo method for the simulation; we repeat a sampling cycle many times enough to obtain a desired confidence interval of the logical error rate per unit simulating time. Each cycle is structured as follows: We first prepare a cluster state described above. Due to the failures of $n$-BSM with the probability of $p_\mathrm{f}$, qubits are randomly removed by the method described in Sec.~\ref{sec:result}. Check operators containing the removed qubits are merged with adjacent check operators repeatedly until not containing any one of them \cite{AAG+18}. If a qubit on a boundary is removed, the involved check operator is removed and the boundary is deformed to include the other qubits in the check operator. If the two opposite $x$-boundaries meet due to the deformation, we conclude that a \textit{logical loss} occurs, namely, that the desired computation fails. In this case, we stop the cycle and start the next one immediately. Otherwise, dephasing errors are randomly assigned to the left qubits with the probability of $p_Z$. For simplicity, qubits on the $t$-boundaries are assumed to be perfect; namely, they are neither removed nor have errors. This ensures that the $t$-boundaries cause no logical losses or errors. Such an unrealistic assumption has negligible effects if $T$ is large enough.

Next, the outcomes of check operators are calculated, then decoded to deduce errors with Edmonds' minimum-weight perfect matching algorithm (MWPM) \cite{edmonds1965paths, edmonds1965maximum, fowler2015minimum} via Blossom V software \cite{kolmogorov2009blossom}.
Error chains connecting the two opposite $x$-boundaries are identified by comparing the assigned and decoded errors.
We then count the number of distinct simulating times corresponding to the ends of the error chains at the boundary of $x=0$, called \textit{erroneous simulating times}.

After repeating enough cycles, we calculate the logical error rate per unit simulating time $p_L$ by the ratio of the number of erroneous simulating times to the total simulating times.
The error threshold $p_\mathrm{th}$ is obtained from the calculated $p_\mathrm{L}$ results for different values of $d$ and $p_Z$; $p_\mathrm{L}$ decreases as $d$ increases if $p_Z < p_\mathrm{th}$ and vice versa otherwise.

\begin{table}
\begin{tabular}{|c |c|c|c|c|c |} \hline
~~ $k$~~ &Possible  & Creation process & Average number \\
        &    GHZ states  &                         &  of $\ket{{\rm GHZ}_3}$\\ \hline
 1 &  $\ket{{\rm GHZ}_4}$&  $\ket{{\rm GHZ}_3}~+_{\rm B_S}~ \ket{{\rm GHZ}_3}$&4 (4.08)\\ \hline
 2 &  $\ket{{\rm GHZ}_5}$&  $\ket{{\rm GHZ}_4}~+_{\rm B_S}~ \ket{{\rm GHZ}_3}$&10 (10.37)\\ 
   &  $\ket{{\rm GHZ}_6}$&  $\ket{{\rm GHZ}_4}~+_{\rm B_S}~ \ket{{\rm GHZ}_4}$&16 (16.66 )\\ \hline
 3 &  $\ket{{\rm GHZ}_7}$&  $\ket{{\rm GHZ}_5}~+_{\rm B_S}~ \ket{{\rm GHZ}_4}$&28 (29.50) \\ 
   &  $\ket{{\rm GHZ}_8}$&  $\ket{{\rm GHZ}_5}~+_{\rm B_S}~ \ket{{\rm GHZ}_5}$& 40 (42.33)\\ 
   &  $\ket{{\rm GHZ}_9}$&  $\ket{{\rm GHZ}_6}~+_{\rm B_S}~ \ket{{\rm GHZ}_5}$& 52 (55.16)\\ 
  &  $\ket{{\rm GHZ}_{10}}$&  $\ket{{\rm GHZ}_6}~+_{\rm B_S}~ \ket{{\rm GHZ}_6}$&64 (68.00)\\ \hline
4 &  $\ket{{\rm GHZ}_{11}}$&  $\ket{{\rm GHZ}_7}~+_{\rm B_S}~ \ket{{\rm GHZ}_6}$&88 (94.19) \\ 
 &  \vdots&  \vdots&\vdots \\ 
 &  $\ket{{\rm GHZ}_{18}}$&  $\ket{{\rm GHZ}_{10}}~+_{\rm B_S}~ \ket{{\rm GHZ}_{10}}$&256 (277.55)\\ \hline
\end{tabular}
\caption{Possible GHZ states at each step $k$ and the corresponding generation processes are tabulated. $+_{\rm B_S}$ stands for the BSM ${\rm B_S}$. 
Average number of  $\ket{{\rm GHZ}_3}$'s consumed in generation of each GHZ state too is tabulated.  The numbers in the brackets in the last column corresponds to the average number of  $\ket{{\rm GHZ}_3}$'s consumed in the presence of photon loss of rate $\eta=0.01$. The photon loss also reduces the success rate of BSM  to $\left(1-2\eta\right)/2$ which in turn increases the average number of $\ket{{\rm GHZ}_3}$'s consumed.} 
\label{tab:ghz}
\end{table}

\section{Counting the $\ket{{\rm GHZ}_3}$'s to generate  $\ket{{\rm GHZ}_r}$}
\label{app:GHZ_r}
The \emph{average} total number of $\ket{\mathrm{GHZ}_3}$'s required to perform one successful BSM of two GHZ states of sizes $m_1$ and $m_2$ is given by $2(N_{m_1}+N_{m_2})$, where $N_{m_1}$, for instance, is the number of $\ket{\mathrm{GHZ}_3}$'s used to generate $\ket{\mathrm{GHZ}_{m_1}}$, and the factor 2 accounts for the 1/2 success rate of a BSM. Using a shorthand notation $+_{\rm B_S}$, we may define $N_{m=m_1+m_2-2}=N_{m_1}+_{\rm B_S}N_{m_2}\equiv2(N_{m_1}+N_{m_2})$, where the resulting $\ket{\mathrm{GHZ}_m}$ from this BSM is always two less than the sum of the constituent sizes. The operation $+_{\rm B_S}$ is non-associative---$A+_{\rm B_S}(B+_{\rm B_S}C)\neq(A+_{\rm B_S}B)+_{\rm B_S}C$.

Based on the above iterative generation protocol, it takes $k=\lceil\log_2(m-2)\rceil$ steps to create a $\ket{\mathrm{GHZ}_m}$ from a minimal set of $M=m-2$ $\ket{\mathrm{GHZ}_3}$'s. The average numbers of $\ket{{\rm GHZ}_3}$'s required to build GHZ states of various sizes are listed in Tab.~\ref{tab:ghz}.
\begin{figure}[t]
\includegraphics[width=0.9\columnwidth]{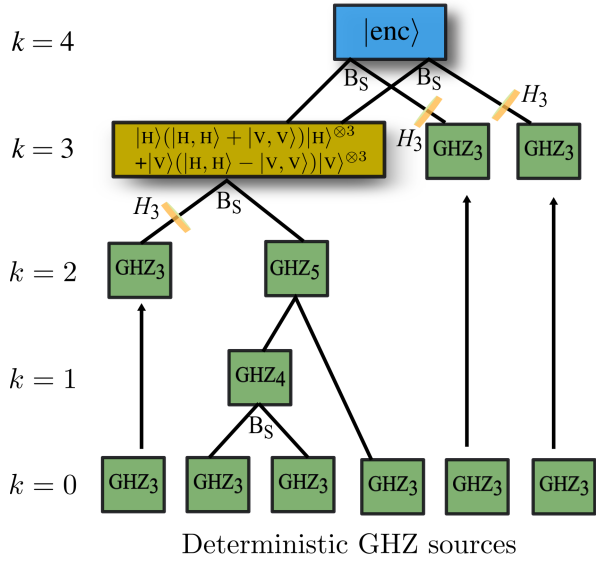}
\caption{  
Schematic representation of the process of generation of $\ket{{\rm enc}}$ using   $\ket{{\rm GHZ}_3}$'s and ${\rm B_S}$ in $k=4$ steps. $H_3$ is the Hadamard operation on the third photon of $\ket{{\rm GHZ}_3}$' and  is performed before  feeding it to ${\rm B_S}$. }
\label{fig:enc}
\end{figure}

According to simple geometric-sum identities, we additionally note that entangling a set of $M$ $\ket{\mathrm{GHZ}_3}$ kets \emph{in any fixed sequential order} yields the resource requirement $N_M=3\cdot2^{M-1}-2$. Such a naive way of entangling GHZ-3 states can result in an excessively large $N_M$. If we define an entangling step as the step in which a maximal number of \emph{independent} BSM are carried out, then an optimal way of entangling GHZ states is to minimize the number of sequential operations at each step.

Starting with the step counter $k=1$ and $m-2$ GHZ-3 states needed to create a $\ket{\mathrm{GHZ}_m}$, an efficient recipe for creating a $\ket{\mathrm{GHZ}_m}$ from BSM of $\ket{\mathrm{GHZ}_3}$'s as basic ingredients can be presented in the following iterative scheme:
\begin{enumerate}
	\item Let $M$ denote the total number of ingredient GHZ states to be entangled using BSMs. When $k=1$, for example, \mbox{$M=m-2$}.
	\item If $M$ is odd, define the number of GHZ pairs $n_\mathrm{p}=(M-1)/2$ with $n_\mathrm{left}=1$ GHZ ket leftover. Otherwise, define $n_\mathrm{p}=M/2$ and $n_\mathrm{left}=1$.
	\item Proceed with $n_\mathrm{p}$ distinct pairwise BSM of GHZ states. If $M$ is odd, $n_\mathrm{left}=1$ GHZ state will not be entangled, which shall be pairwise entangled with another GHZ state in the next step.
	\item If $M=1$, terminate the generation protocol. Otherwise, update $M=n_\mathrm{p}+n_\mathrm{left}$ and raise $k$ by one.
\end{enumerate}

Using this numerical recipe, the (unbracketed) values in Tab.~\ref{tab:ghz} can be generated. Upon numerical-pattern inspection, we find the explicit analytical formula,
\begin{equation}
    N_m=3(m-2)\cdot 2^{\lfloor\log_2(m-2)\rfloor}-2\cdot 4^{\lfloor\log_2(m-2)\rfloor}\,,
    \label{eq:num_GHZ3_for_GHZm}
\end{equation}
for the number of GHZ$_3$ states needed to generate a GHZ$_m$ state.

\section{Generation of concatenated resource state}
\label{app:newc3}
In this section we describe how to create resource state concatenated with three-qubit repetition QEC code that is, $\ket{\mathcal{C}_{3^\prime}}_{\rm enc}=\ket{0_{n}}\left(\ket{0_m}+\ket{1_m}\right)^{\otimes3}\ket{0_{n}}+\ket{1_{n}}\left(\ket{0_m}-\ket{1_m}\right)^{\otimes3}\ket{1_{n}}$. This state can be generated in the $[k=\log_2(m-1)]$th step when using $\ket{\rm{GHZ}_3}$. To begin with, apply Hadamard operation on $\ket{{\rm GHZ_{m+1}}}$ so that we have $H_{m+1}\ket{{\rm GHZ_{m+1}}}=\left(\ket{\textsc{h}}^{\otimes m}+\ket{\textsc{v}}^{\otimes m}\right)\ket{\textsc{h}}+\left(\ket{\textsc{h}}^{\otimes m}-\ket{\textsc{v}}^{\otimes m}\right)\ket{\textsc{v}}$. Applying ${\rm B_S}$ between  $H_{m+1}\ket{{\rm GHZ_{m+1}}}$ and  $\ket{{\rm GHZ_{5}}}$, and rearranging the modes we get $\ket{\textsc{h}}\left(\ket{\textsc{h}}^{\otimes m}+\ket{\textsc{v}}^{\otimes m}\right)\ket{\textsc{h}}^{\otimes3}+\ket{\textsc{v}}\left(\ket{\textsc{h}}^{\otimes m}-\ket{\textsc{v}}^{\otimes m}\right)\ket{\textsc{v}}^{\otimes3}$ in ($k+1$)th step.
Further, by entangling two $H_{m+1}\ket{{\rm GHZ_{m+1}}}$'s with the above state the encoded central qubit,
\begin{align}
\ket{\rm{enc}}=&\,\,\ket{\textsc{h}}\left(\ket{\textsc{h}}^{\otimes m}+\ket{\textsc{v}}^{\otimes m}\right)^{\otimes3}\ket{\textsc{h}}\nonumber\\
&+\ket{\textsc{v}}\left(\ket{\textsc{h}}^{\otimes m}-\ket{\textsc{v}}^{\otimes m}\right)^{\otimes3}\ket{\textsc{v}}\,,
\end{align} 
in $k+2$-th step. Finally, two $\ket{{\rm GHZ_{n+1}}}$'s are entangled on both sides of $\ket{\rm{enc}}$ to get the desired state, $\ket{\mathcal{C}_{3^\prime}}_{\rm enc}$. This final step is nothing but replacing $\ket{\rm{GHZ}_{m+2}}$ with $\ket{\rm{enc}}$  in the $k$th step in Fig.~\ref{fig:c3}.
In the step $k=2$, $(1+10)/0.5=22$ $\ket{\rm{GHZ}_{3}}$'s are consumed on average. Further,  $(22+2)/(0.5\times0.5)=96$ $\ket{\rm{GHZ}_{3}}$'s are consumed  on average in creation of a $\ket{\rm{enc}}$. However, taking in to account the photon loss of rate $\eta=0.01$, the average number of $\ket{\rm{GHZ}_{3}}$'s consumed in ($k=2$)th step is $(1+10.37)/[(1-2\eta)/2]=23.20$  and finally it is  $(23.20+2)/[(1-2\eta)/2]^2\approx104.96$. Figure~\ref{fig:enc} supplements these statements with a flowchart.

When we have $m=2$, The first process in creation of $\ket{\rm{enc}}$ takes place in $k=2$ time steps. Therefore, creation of $\ket{\rm{enc}}$ takes a total of 4 steps. In the unencoded case, the central qubit always waited for 2 time steps during creation of $\ket{\mathcal{C}_{3^\prime}}$ and totally $k=4~(5)$  for $\ket{\mathcal{C}_{\mathcal{L}}}$ in MTQC-1~(MTQC-2). In the encoded case the central qubit has to wait for totally $k=6~(7)$ time steps before being measured for quantum computing.

\end{document}